\documentclass[superscriptaddress,secnumarabic,prd,aps,showpacs,nofootinbib,showkeywords,eqsecnum]{revtex4-1}

\usepackage{graphicx,color,amsmath,amsxtra}
\usepackage{epsf}
\usepackage{amssymb}
\usepackage{enumerate}
\usepackage{hhline}
\usepackage{array}
\usepackage{tabularx}
\usepackage[unicode]{hyperref}

\begin{document}
\pagestyle{myheadings}

\title{CMB imprints of non-canonical anisotropic inflation}
\author{Tuan Q. Do}
\email{tuan.doquoc@phenikaa-uni.edu.vn}
\affiliation{Phenikaa Institute for Advanced Study, Phenikaa University, Hanoi 12116, Vietnam}
\affiliation{Faculty of Basic Sciences, Phenikaa University, Hanoi 12116, Vietnam}
\affiliation{Faculty of Physics, VNU University of Science, Vietnam National University, Hanoi 11416, Vietnam}
\author{W. F. Kao}
\email{gore@mail.nctu.edu.tw}
\affiliation{Institute of Physics, National Chiao Tung University, Hsin Chu 30010, Taiwan}
\author{Ing-Chen Lin}
\email{g9522528@oz.nthu.edu.tw}
\affiliation{Institute of Physics, National Chiao Tung University, Hsin Chu 30010, Taiwan}
\date{\today} 

\begin{abstract}
Effect of non-canonical scalar fields on the CMB imprints of the anisotropic inflation will be discussed in details in this paper. In particular, we are able to obtain the general formalism of the angular power spectra in the scalar perturbations, tensor perturbations, cross-correlations, and linear polarization in the context of the anisotropic inflation model involving non-canonical scalar fields. Furthermore, some significant numerical spectra will be plotted using the most recent data of Planck as well as the BICEP2 and Keck array.  As a result, we find a very interesting point that the $TT$ spectra induced by the tensor perturbations as well as by the  linear polarization will increase when the speed of sound decreases.

\end{abstract}

%


\maketitle
\section{Introduction} \label{intro}
Recently, some anomalies of the cosmic microwave background radiation (CMB) such as the hemispherical asymmetry and the cold spot have been detected by the Wilkinson Microwave Anisotropy Probe (WMAP) satellite \cite{Komatsu:2010fb} and Planck one \cite{Planck,Planck2}. 
These phenomena are beyond the prediction of standard inflationary models \cite{guth}. Consequently, there have been a number of mechanisms proposed to explain the origin of these anomalies, which can be found in a recent interesting review paper \cite{Schwarz:2015cma}.
A possibility for this inconsistency might be derived from the  assumption  that the spacetime of early universe is the homogeneous and isotropic Friedmann-Lemaitre-Robertson-Walker (FLRW) metric \cite{FLRW}. 
Hence, the effect of cosmic inflation in anisotropic Bianchi type space time \cite{bianchi} could lead to a proper resolution providing corrections consistent with observations \cite{Pitrou:2008gk}.  It is worth noting that some studies, e.g., those in Refs. \cite{Hanson:2010gu,Bennett:2012zja,data1}, have claimed that the CMB statistical anisotropy could be instrumental rather than cosmological. In particular, these investigations using either the WMAP or Planck data have pointed out that the asymmetric beams could cause the CMB statistical anisotropy. However, this explanation has been tested independently by other people in Ref. \cite{data}. As a result, they have found that  the asymmetric beams seem to be unimportant \cite{data}.

If the early universe was induced by anisotropic inflation, we would like to find out the late-time state of universe. There is an important hint from the cosmic no-hair conjecture proposed by Hawking and his colleagues predicting that the universe will approach a homogeneous and isotropic state at late time regardless of any anisotropic and/or inhomogeneous initial states \cite{GH}. 
If the conjecture is true, the universe would be isotropic and homogeneous at late time. 
Unfortunately, the conjecture is difficult to prove. 
Indeed, there have been efforts trying to prove the cosmic no-hair conjecture for many different models. 
For example, Wald has proposed a proof using the energy conditions approach for the Bianchi metrics, which are homogeneous but anisotropic spacetime \cite{wald}.
Kleban {\it et al.} have also tried to generalize the Wald's proof to anisotropic and inhomogeneous spacetimes recently \cite{inhomogeneous}. 
More interestingly, Carroll and Chatwin-Davies have proposed another proof for the conjecture using the entropy approach \cite{Carroll:2017kjo}. 
In addition, some proofs for specific inflation models have been carried out in Ref. \cite{Barrow:1987ia}. 
Furthermore, some observational constraints for anisotropic universe have been analyzed in Ref. \cite{Saadeh:2016sak}, which could be useful in order to see whether the proposed anisotropic inflation models are viable.

It appears that the Wald's proof is incomplete since it is no longer valid for inhomogeneous spacetimes. Independently, Starobinsky showed in a seminal paper  \cite{Starobinsky:1982mr} that an inflationary solution of the Einstein gravity in the presence of cosmological constant $\Lambda$ will approach globally to an isotropic but inhomogeneous state at late time. Furthermore, by regarding primordial scalar and tensor perturbations as the hairs he pointed out that the inhomogeneous and time-independent tensor hairs remain outside of the future event horizon of an observer. This result indicates that if the inflationary universe turned out to be the isotropic de Sitter state at late time, it would be local, i.e., inside of the future event horizon. Additionally,  a power-law inflation scenario was then considered to examine the validity of this interesting result \cite{Muller:1989rp}. As a result, it was shown that a global cosmic no-hair conjecture would not exist in this model as expected. All these works imply that the Hawking cosmic no-hair conjecture could only be local, i.e., valid inside of the future event horizon. Therefore, it could not predict a global asymptotic structure of inflationary spacetime outside of the event horizon.  Along with the Starobinsky's studies, other people also investigated inhomogeneous background spacetimes, e.g., the so-called Tolman-Bondi spacetime, through different approaches \cite{Barrow:1984zz}. In particular, scale factors of background spacetimes in these models are regarded as functions of not only time coordinate but also spatial coordinate(s). In addition, the approach using the small perturbations of de Sitter spacetime \cite{Starobinsky:1982mr,Muller:1989rp}, which turns out to be technically difficult, was not chosen in these works. Instead, an effective approach, in which the background spacetimes are initially assumed to be anisotropic and/or inhomogeneous  \cite{Barrow:1984zz}, was considered. As a result, all these works arrived  at the same conclusion that the late time state of the universe would be locally isotropic and homogeneous.

Besides the proofs, it has been shown that the cosmic no-hair conjecture might be violated in some particular gravity models \cite{barrow06,kao09,kaloper,MW}. 
For example, the cosmic no-hair conjecture was claimed to break down for the higher-order extensions of general relativity in Ref. \cite{barrow06}.
It has also been pointed out in Ref. \cite{kaloper} that the Lorentz Chern-Simons terms might disfavor the prediction of this conjecture. 
However, some of these counterexamples have been shown to be unstable by stability analysis indicating that the cosmic no-hair conjecture could be true for these models \cite{kao09}. This result indicates that the Einstein gravity with or without corrections might not maintain any classical hair of early universe. 
However, hairs could exist for models with matter fields with special coupling shown in Ref. \cite{MW}.
For example, a supergravity-motivated model with a unusual coupling term between scalar and vector fields, $f^2(\phi)F_{\mu\nu}F^{\mu\nu}$, has been proposed by Kanno, Soda, and Watanabe (KSW) \cite{MW}. 

Indeed, this model has been shown to admit a stable and attractor Bianchi type I solution during an inflationary phase.
The new set of solutions has been shown to disfavor the cosmic no-hair conjecture. 
Furthermore, this result is still valid when a canonical scalar field of the KSW model is replaced by non-canonical models such as the Dirac-Born-Infeld (DBI) field \cite{WFK1}, the supersymmetric Dirac-Born-Infeld (SDBI) field \cite{Do:2016ofi}, the $k$-inflation \cite{Ohashi:2013pca}, and the convariant Galileon field \cite{WFK2}. 
More interestingly, the cosmic no-hair conjecture has been shown to be broken down in a five-dimensional extension of the KSW model \cite{WFK3}. 
As a result, the unusual coupling $f^2(\phi)F_{\mu\nu}F^{\mu\nu}$ has played an interesting role inducing a stable anisotropy during the inflationary phase to a spacetime of the KSW model as well as its non-canonical extensions. 
Many other extensions of the KSW model have also been proposed in Ref. \cite{extensions} to check if the cosmic-no-hair conjecture is valid or not. 

In summary, the KSW model has become an attractive model \cite{SD}. 
Hence, it is very interesting to study all cosmological aspects of this anisotropic inflation model. 
Some significant deviations from predictions of isotropic inflation could be from these models.
The predictions can also be checked by more sensitive observations in the near future. 
For example, imprints of the KSW anisotropic inflation on the CMB through correlations between $T$, $E$, and $B$ modes have been investigated systematically in Refs. \cite{Imprint,Imprint1,Imprint2,Imprint3}, while primordial gravitational waves for the KSW model have been discussed in Ref. \cite{gws}. 
The primordial gravitational waves detectors like the BICEP2 or Keck array \cite{Ade:2018gkx} will also provide critical constraints for these theoretical predictions.
In this paper, therefore, we propose to study the correlations of $T$, $E$, and $B$ modes in non-canonical anisotropic inflation, e.g., the DBI and SDBI models \cite{WFK1,Do:2016ofi} to see how the non-canonical terms of scalar field affect the results in the KSW model \cite{Imprint,Imprint1,Imprint2}. 

As a result, this paper will be organized as follows: (i) An introduction to this study has been written in the Sec. \ref{intro}. (ii) A short setup of the non-canonical anisotropic inflation models will be shown in Sec. \ref{sec2}. (iii) Then, scalar and tensor perturbations for these anisotropic inflation models will be presented in Sec. \ref{sec3}. (iv) Accordingly, the CMB imprints of the non-canonical anisotropic inflation will be shown analytically in Sec. \ref{sec4}. In addition, numerical spectra will be plotted using the recent data of Planck as well as the BICEP2 and Keck array in this section. (v) Finally, concluding remarks will be given in Sec. \ref{final}. 

\section{Non-canonical extensions of Kanno-Soda-Watanabe model} \label{sec2}
In principle, a non-canonical extension of Kanno-Soda-Watanabe model can be built by replacing canonical scalar field by non-canonical model. 
For example, a non-canonical extension of Kanno-Soda-Watanabe model can be described by the following action \cite{Ohashi:2013pca}
\begin{equation}\label{general lagra}
S = \int {d^4 } x \sqrt {g} \left[ {\frac{{M_p^2 }}
{2} R+ P(\phi,X) - \frac{1}
{4}h^2 \left( \phi  \right)F_{\mu \nu } F^{\mu \nu }  } \right],
\end{equation}
with $M_p$ the reduced Planck mass, $F_{\mu \nu }  \equiv \partial _\mu  A_\nu   - \partial _\nu  A_\mu  $ the field strength of the vector field $A_\mu$, and $h(\phi)$ an arbitrary function of scalar field respectively \cite{MW,SD,WFK1,Do:2016ofi,Ohashi:2013pca,WFK2,WFK3,extensions,Imprint,Imprint1,Imprint2,Imprint3,gws}.  
Note that $P(\phi,X)$ is an arbitrary function of $\phi$ with $X\equiv -\partial^\mu \phi \partial_\mu \phi/2$ the standard kinetic term \cite{k-inflation}. 
It is clear that if $P(\phi,X)$ is of the following form,
\begin{equation}
P(\phi,X) = \frac{1}
{{f\left( \phi  \right)}}\frac{\gamma-1}{\gamma} -V\left( \phi  \right) ,
\end{equation}
then we will have the corresponding DBI extension of the KSW model \cite{WFK1}. Here  $\gamma  \equiv 1/ \sqrt {1 + f (\phi) \partial _\mu  \phi \partial ^\mu  \phi}$ is the Lorentz factor characterizing the motion of the D3-brane \cite{WFK1}. 
It is clear that $\gamma \geq 1$ for non-negative $f (\phi)$. 
On the other hand, if $P(\phi,X)$ is chosen to be
\begin{equation}
P(\phi,X) =\frac{1}
{{f\left( \phi  \right)}}\frac{\gamma-1}{\gamma} -\Sigma_0^2 V\left( \phi  \right) ~\text{with}~\Sigma_0=\left({\frac{\gamma+1}{2\gamma}}\right)^{\frac{1}{3}}
\end{equation}
the model reduces to the supersymmetric DBI extension of the KSW model \cite{Do:2016ofi}. 
Additionally, another non-canonical extension of the KSW model can be found in Ref. \cite{Ohashi:2013pca} with $P(\phi,X)$ assumed to be the generalized ghost condensate form
\begin{equation}\label{ghost cond}
 P(\phi,X) = -X + \frac{c}{M_p^{4n}} \exp \left[\frac{n\lambda\phi}{M_p}\right] X^{n+1}.
\end{equation}
Here $c>0$ and $n \geq1$ are all constants \cite{Ohashi:2013pca}. 
Note that we can also go beyond the non-canonical action shown in Eq. (\ref{general lagra}) by adding more non-canonical terms of scalar field such as, $G(\phi,X) \square \phi$ instead of the function $P(\phi,X)$. 
These extensions are referred to as the (covariant) Galileon extension of the KSW model \cite{WFK2}. 
We will focus, however, on the model \eqref{general lagra} in this paper. 
\section{Scalar and tensor perturbations} \label{sec3}
Note that the anisotropy deviation $\sigma$ should be very small compared to the isotropy parameter $\alpha$ in order to be consistent with the observations data like  WMAP  \cite{Komatsu:2010fb} or  Planck \cite{Planck2}. 
Hence, it is a good approximation to take the background metric as the spatially flat Friedmann-Lemaitre-Robertson-Walker (FLRW) metric instead of the Bianchi type I metric for simplicity \cite{Imprint,Imprint1,Imprint2,Imprint3}.  
We will also present a brief review of the derivation shown in Refs. \cite{Imprint,Imprint1,Imprint2,Imprint3}.

\subsection{Scalar perturbation}
In this subsection, we will briefly review some results on the scalar perturbation presented in Ref. \cite{Do:2016ofi}. 
Following Refs. \cite{k-inflation,perturb}, the scalar perturbation metric can be written as
\begin{equation} \label{longitudinal gauge}
ds^2 = a^2(\eta) \left[ -\left(1+2\Phi\right)d\eta^2 +\left(1+2\Psi \right)\delta_{ij}  dx^{i} dx^j\right]
\end{equation}
with $\eta \equiv \int {a^{-1}dt}$ the conformal time.
The gauge covariant curvature perturbation is given by \cite{k-inflation}
\begin{equation}
 \zeta = \Psi - \frac{H}{\dot\phi} \delta \phi,
\end{equation}
with $H$ the Hubble constant.
In particular, the conformal Newtonian (or longitudinal) gauge with $\Psi=-\Phi$ has been adopted in deriving the scalar perturbations for the power spectrum in the literature \cite{k-inflation,perturb}. 
On the other hand, an alternative spatially flat gauge with $\Psi=0$ 
\begin{equation} \label{flat-gauge}
 \zeta = - \frac{H}{\dot\phi} \delta \phi
\end{equation} 
has also been used in the literature \cite{Imprint,Imprint1,Imprint2,Imprint3}. Both gauge choices will lead effectively to the same results in the derivation of perturbations of power spectrum.
This is due to the fact that metric sector perturbations can be shown to be either slow-roll suppressed or canceled with each other as compared to the matter perturbations in deriving the perturbed power spectrum \cite{Emami:2013bk,Chen:2005fe,CHKS}.
Indeed, it can be shown that \cite{k-inflation}
\begin{equation}
\zeta =-\Phi - {2 \rho \over 3(\rho +p)} \left( \Phi + {\dot \Phi \over H} \right), 
\end{equation}
with $\rho$ and $p$ the energy density and pressure density associated with the scalar field energy momentum tensor. 
Hence $\zeta \simeq -{2 \rho  \Phi} / 3(\rho +p)$ in the slow-roll limit ($ \rho  \gg \rho +p$) for the non-decaying mode ($\dot \Phi \ll  H$). 
This proves that the contribution of the scalar curvature term $\Phi$ is indeed slow-roll suppressed as compared to the scalar field contribution in $\zeta$ \cite{CHKS}. 

For the action (\ref{general lagra}),
it can be shown in Refs. \cite{MW,Imprint,Imprint1} that the existence of a vector field coupled to scalar field could induce a small spatial anisotropy. 
Indeed, once the statistical isotropy of CMB is broken, the scalar power spectrum becomes \cite{ACW}
\begin{equation} \label{ACW formula}
{\cal P}^\zeta_{k,{\text {ani}}}  = {\cal P}^{\zeta(0)}_{k} \left(1+g_\ast \cos^2 \theta_{{\bf k},{\bf V}}\right) .
\end{equation} 
Here $g_\ast$ is a constant characterizing the deviation from the spatial isotropic and is expected to be smaller than one, i.e., $|g_\ast| <1$.
In addition,  $\theta_{{\bf k},{\bf V}}$ is the angle between the comoving wave number $\bf k$ with the {\it privileged} direction $\bf V$  close to the ecliptic poles \cite{ACW}. 
Moreover, we will write $\theta_{{\bf k},{\bf V}}$ as $\theta$ for convenience.

It is worth noting that the non-vanishing of $g_\ast$ has been observed \cite{data,data1,data2,data3,data4}. 
In particular,  it was shown \cite{data} that $g_\ast =0.29 \pm 0.031$, nonzero at $9\sigma$ using the 5-year  WMAP data.
In addition, the bound $g_\ast = 0.002 \pm 0.016$  has been obtained at 68\% confidence level using the Planck 2013  data \cite{data1}. 
Moreover, the other bound  $|g_\ast|<0.072$  has also been obtained at 95\% confidence level ($-0.046< g_\ast < 0.048$ at 68 \% confidence level) using the 9-year WMAP data\cite{data2}.
A more general analysis using the Planck 2015 data for $g_\ast$ has been carried out \cite{data3} to provide the bounds that $-0.041<g_\ast <0.036$ at 95\% confidence level. 
Another analysis using three dimensional spectroscopic galaxy data of LSS surveys has revealed that $-0.09<g_\ast < 0.08$ at  95\% confidence level \cite{data4}. 

In addition, the isotropic scalar power spectrum ${\cal P}^{\zeta(0)}_{k} $ is given by \cite{k-inflation}
\begin{eqnarray}  \label{def of P-zeta-2}
{\cal P}^{\zeta(0)}_{k}  =  \left. \frac{1}{8\pi^2 M_p^2}\frac{H^2}{c_s \epsilon}\right|_{c_s^\ast k_\ast = a_\ast H_\ast},
\end{eqnarray}
with a general definition for non-canonical scalar field.
Here $\epsilon\equiv -\dot H/H^2$ denotes the slow-roll parameter, $H\equiv \dot a/a$ is the Hubble expansion rate, and $a(t)$ is a scale factor of metric. 
In addition, ${^\ast}$ denotes the pivot scale (or horizon-exit scale) where the universe can be approximated by de Sitter space specified by $a_\ast \simeq -(H_\ast \eta_\ast)^{-1}$. 
Note also that the notation $c_s$ stands for the speed of sound defined by \cite{k-inflation}
\begin{equation} \label{speed of sound}
c_s^2 \equiv \frac{\partial_X p}{\partial_X \rho}=\frac{\rho+p}{2X \partial_X \rho},
\end{equation}
with $p$ and $\rho$  the pressure and energy density parameters defined by
\begin{eqnarray}
p = &&  ~ P(\phi,X),\\
\rho = && ~ 2X \partial_X P(\phi,X)-P(\phi,X),
\end{eqnarray}
respectively. 
It is straightforward to show that $c_s =1$ for canonical scalar field and $c_s^2 \neq 1$ for non-canonical scalar fields. 
Indeed, it turns out that  
\begin{equation}
c_s^2=\frac{1}{\gamma^{2}}, 
\end{equation}
for the DBI field.
In addition, $c_s^2$ for the SDBI field is given by \cite{Do:2016ofi}
\begin{equation}
c_s^2 =\frac{\Sigma_2}{\Sigma_1}, 
\end{equation}
with
\begin{eqnarray}
&&  \label{sig1}
\Sigma_1 = \gamma^3 + \frac{\gamma}{9\Sigma_0} \left({3\gamma^2+\gamma-1}\right)f V , \\
&& \label{sig2}
\Sigma_2 = \gamma + \frac{\gamma}{3\Sigma_0} f V .
\end{eqnarray}
For the generalized ghost condensate model \eqref{ghost cond},  $c_s$ can be shown to be  \cite{Ohashi:2013pca}
\begin{equation}
c_s^2 =\frac{-1 + (n+1)\frac{c}{M_p^{4n}} \exp \left[\frac{n\lambda\phi}{M_p}\right] X^{n}}{-1 + (n+1)(2n+1)\frac{c}{M_p^{4n}} \exp \left[\frac{n\lambda\phi}{M_p}\right] X^{n}}.
\end{equation}
It is apparent that $\Sigma_2 < \Sigma_1$ and $\gamma \ge 1$ for non-negative $f(\phi)$ and $V(\phi)$. 
Hence, $c_s^2 \leq 1$  for both the DBI and SDBI models. 
Note that $\gamma \simeq 1$ as shown in Ref. \cite{Do:2016ofi} in order to induce inflation. 
Hence, $c_s^2 \simeq \gamma^{-2} \simeq 1$  since $\Sigma_1 \simeq \gamma^2 \Sigma_2 $  in the framework of the SDBI inflation. 
The speed of sound $c_s$  is also smaller than one for $n \geq 1$ for the generalized ghost condensate model \eqref{ghost cond} \cite{Ohashi:2013pca}.

Following Ref. \cite{Imprint}, $g_\ast$ can be evaluated for the non-canonical scalar field. 
As a result, we need only to restore the speed of sound, $c_s^2$, of non-canonical scalar field to appropriate order.  
Note first that the standard Bunch-Davies (BD) vacuum state  for the non-canonical scalar field can be shown to be \cite{CHKS} 
\begin{equation} \label{general Bunch-Davies vacuum}
 \zeta^{(0)}_{\text{nc}} (k,\eta) = \frac{H }{2\sqrt{c_s \epsilon}M_p k^{3/2}} \left(1+i c_s k \eta\right) e^{-i  c_s k \eta} .
\end{equation}
Here the superscript $(0)$ denotes the de Sitter background.
Given this BD vacuum we are able to estimate the full scalar power spectrum  for the non-canonical scalar field. 
Indeed, the full power spectrum in the Heisenberg interaction picture for the scalar perturbation, up to the second order, can be shown to be  \cite{Imprint,Imprint1}
\begin{eqnarray}\label{scalar-power-spectra-1}
\langle 0| \hat{\zeta}_{\text{nc}} (k,\eta) \hat{\zeta}_{\text{nc}} (k',\eta) |0\rangle &\simeq & 
\frac{2\pi^2}{k^3}\delta^3({\bf k}+{\bf k}') {\cal P}^{\zeta(0)}_{k,{\text{nc}}}  - \int^\eta_{\eta_{min,1}}d\eta_1 \int^{\eta_1}_{\eta_{min,2}}d\eta_2\langle 0|  \left[ \left[\hat{\zeta}^{(0)}_{\text{nc}}(k,\eta) \hat{\zeta}^{(0)}_{\text{nc}}(k',\eta), {H}_{\zeta} (\eta_1)\right],{H}_{\zeta} (\eta_2) \right]|0\rangle . \nonumber\\
\end{eqnarray}
Here $ {\cal P}^{\zeta(0)}_{k,{\text{nc}}}$ is  the isotropic scalar power spectrum for the non-canonical scalar field,  defined in Eq. (\ref{def of P-zeta-2}), given by 
\begin{equation} \label{def.of.Pzeta}
 {\cal P}^{\zeta(0)}_{k,{\text{nc}}}={\cal P}^{\zeta(0)}_{k} =\frac{1}{8\pi^2 M_p^2}\frac{H^2}{c_s \epsilon}
\end{equation}
with $\hat{\zeta}^{(0)}_{\text{nc}}(k,\eta) $ defined as
\begin{eqnarray}
\hat{\zeta}^{(0)}_{\text{nc}}(k,\eta) =  \zeta^{(0)}_{\text{nc}} (k,\eta)  a({\bf k}) + \zeta^{(0)\ast}_{\text{nc}} (k,\eta)  a^\dagger (-{\bf k}).
\end{eqnarray}
Here, $a({\bf k})$ and $a^\dagger (-{\bf k})$ are the creation and annihilation operators satisfying the commutation relation $[a({\bf k}), a^\dagger(-{\bf k}') ]=\delta^3({\bf k}+{\bf k}')$, $[a({\bf k}), a ({\bf k})]=0$, and $ [a^\dagger({\bf k}), a^\dagger ({\bf k})]=0$.  
It is hence straightforward to show that the first order term in Eq. (\ref{scalar-power-spectra-1}) vanished due to  $ \langle 0|a^\dagger|0\rangle =0 $ and $ a|0\rangle =0 $. 
Note that ${H}_{\zeta}(\eta)$, in Eq. (\ref{scalar-power-spectra-1}), is the interaction Hamiltonian associated with the scalar perturbations derived from the coupling term $-1/4 h^2(\phi)F^2$ \cite{Imprint,Imprint1}. 
As a result, the scalar perturbation part of the tree-level interacting Lagrangian can be decomposed as $A_i({\bf x},\eta) = A_i^{(0)}(\eta)+\delta A_i({\bf x},\eta) $ ($i=x,~y,~z$) with the field perturbations  \cite{Imprint,Imprint1}
\begin{equation}
L_{{\rm int}, \zeta}=4 a^4 E_x \delta E_x  \zeta^{(0)}_{\text{nc}}.
\end{equation}
Here $E_x\equiv (h/a^2) {A_x^{(0)'}}$ and $a\equiv e^{\alpha}=-(\eta H)^{-1}$ is the scale factor of de Sitter background spacetime.
In addition, $A_x^{(0)}$ is the background value of the $x$-component of the vector field defined by $A_\mu^{(0)} =(0,A_x^{(0)},0,0)$ with ${A_x^{(0)'}} \equiv d {A_x^{(0)}}/d \eta$ ~\cite{Imprint}. 
Moreover, $ \delta { E}_i(\eta, {\bf x}) \equiv h \delta A_i'/a^2 $ are the perturbations electric components. 
Consequently, the corresponding Hamiltonian for the scalar perturbation can be shown to be
\begin{equation}
{H}_{\zeta}(\eta) =4a^4 {E_x}\int d^3k \delta {\cal E}_x({\bf k},\eta)\hat{\zeta}( -{\bf k},\eta),
\end{equation} 
with  $\delta {\cal E}_x({\bf k},\eta)$ the Fourier transform of $\delta { E}_x({\bf x},\eta)$ defined, in the super-Hubble regime with $|k\eta|\ll 1$, as \cite{Imprint}
\begin{equation}
\delta {\cal E}_x( {\bf k},\eta) =\sum_{\lambda=1,2 }^{} \frac{3H^2}{\sqrt{2k^3} } \left[a_\lambda({\bf k})+ a^\dagger_\lambda({\bf -k})   \right] \epsilon_x^{(\lambda)}({\bf k}).
\end{equation}
A more general definition of $\delta {\cal E}_i( {\bf k},\eta)$ is given by
\begin{equation} \label{def.of.delta.cal.E}
\delta {\cal E}_i( {\bf k},\eta) =\sum_{\lambda=1,2 }^{} \frac{3H^2}{\sqrt{2k^3} } \left[a_\lambda({\bf k})+ a^\dagger_\lambda({\bf -k})   \right] \epsilon_i^{(\lambda)}({\bf k}), 
\end{equation}
with $i=x,~y,~z$.
Here $\epsilon_x^{(\lambda)}({\bf k})$ is the $x$-component of the polarization vectors satisfying the following conditions \cite{Imprint}
\begin{equation}
k_i \epsilon_i^{\lambda }({\bf k})=0;~  \epsilon_i^{\lambda }({\bf -k})=\epsilon_i^{\lambda \ast}({\bf k}); ~ \epsilon_i^{\lambda }({\bf k})\epsilon_i^{\lambda' \ast }({\bf k})=\delta_{\lambda\lambda' }.
\end{equation}
Due to  the rotational symmetry in the $(y,z)$ plane, the wave number vector in the Fourier space can be defined as ${\bf k}=k(\cos \theta,\sin \theta,0)$ with $\theta$  the angle between $\bf k$ and $x$-axis. 
Hence, the polarization vectors are defined as
\begin{equation}
\epsilon_i^{1 }({\bf k})=(-i\sin\theta , i \cos\theta, 0);~ \epsilon_i^{2 }({\bf k})=(0,0,1) . \label{ep12}
\end{equation}
As a result, we can show that the last term in Eq. (\ref{scalar-power-spectra-1}) reduces to
\begin{equation} \label{scalar-power-spectra-2}
-\int^\eta_{\eta_{min,1}}d\eta_1 \int^{\eta_1}_{\eta_{min,2}}d\eta_2\langle 0| \left[ \left[\hat{\zeta}^{(0)}_{\text{nc}}(k,\eta) \hat{\zeta}^{(0)}_{\text{nc}}(k',\eta), {H}_{\zeta} (\eta_1) \right],{H}_{\zeta} (\eta_2) \right]|0\rangle =\frac{2\pi^2}{k^3}\delta^3({\bf k}+{\bf k}')   \frac{c_s^4 E_x^2 N_{c_sk}^2 }{\pi^2 \epsilon^2 M_p^4} \sin^2 \theta.
\end{equation}
Here we have used the fact that $\eta_{min,i}=-(c_s k_i)^{-1} $ at the horizon exit. A brief review is presented in Appendix  \ref{app1}. 
Hence, the full power spectrum becomes
\begin{eqnarray} \label{general anisotropic scalar power spectrum}
{\cal P}^\zeta_{k,{\text{nc}}}  
 = && ~ {\cal P}^{\zeta(0)}_{k,{\text{nc}}} \left(1+ \frac{8c_s^5 E_x^2  N_{c_sk}^2}{\epsilon M_p^2 H^2}  \sin^2 \theta \right) ,
\end{eqnarray}
with  $N_{c_sk} \simeq 60 $ the e-fold number.  
As a result, $ g_\ast$ for non-canonical scalar field can be defined by comparing Eq. (\ref{general anisotropic scalar power spectrum}) with Eq. (\ref{ACW formula}) as
\begin{equation} \label{definition_of_g}
 g_{\ast } = - c_s^5 \frac{8 E_x^2  N_{c_sk}^2}{\epsilon M_p^2 H^2} = c_s^5 g_\ast^0 <0.
\end{equation}
Here 
\begin{equation} \label{definition_of_g0}
g_\ast^0 = - \frac{8 E_x^2  N_{c_sk}^2}{\epsilon M_p^2 H^2} <0
\end{equation}
for the canonical scalar field \cite{Imprint}. 

 It is apparent that $ g_\ast \simeq g_\ast^0$ if  $c_s^2 \simeq 1$ (e.g., in the SDBI model with $\gamma \simeq 1$). 
On the other hand, $|g_\ast| \ll |g_\ast^0|$ if  $c_s^2 \ll 1$. 
Consequently,  ${\cal P}^\zeta_{k,{\text{nc}}} \sim {\cal P}^{\zeta(0)}_{k,{\text{nc}}}$ for the models with $c_s^2 \ll 1$ (e.g., $\gamma \gg 1$ for DBI model).

Note also that the scalar spectral index can be shown to be 
\begin{eqnarray} \label{general scalar spectral index}
 n_s - 1 \equiv && ~ \left. \frac{d \ln {\cal P}^\zeta_{k,{\text{nc}}}   }{d \ln k} \right|_{c^{\ast}_s k_\ast = a_\ast H_\ast} \nonumber\\
\simeq && ~ -2\epsilon - \tilde \eta -s +\left(\frac{2}{N_{c_sk}}-5s\right) \frac{ g_\ast \sin^2 \theta}{1-g_\ast \sin^2 \theta }, 
\end{eqnarray}
where  $\tilde \eta \equiv \dot\epsilon/(\epsilon H)$, $s\equiv \dot c_s/(c_s H)$ \cite{k-inflation,Baumann}.
We have used the result that  $d r_A/dt \simeq 0$ during the inflationary phase \cite{Imprint}. 
With the average value of  $\sin^2 \theta$ given by $\langle \sin^2 \theta \rangle =2/3$ \cite{Imprint}, Eq. (\ref{general scalar spectral index}) reduces to
\begin{eqnarray} \label{general scalar spectral index 1}
 n_s - 1 \simeq  -2\epsilon - \tilde \eta -s +\left(\frac{2}{N_{c_sk}}-5s\right) \frac{2  g_\ast }{3-2g_\ast  }.
\end{eqnarray}
\subsection{Tensor perturbations}
The tensor perturbation metric can be written as \cite{Imprint,perturb}
\begin{equation}
g_{\mu\nu}=a^2(\eta) \left[-d\eta^2 + \left(\delta_{ij}+h_{ij}\right)dx^i dx^j\right],
\end{equation}
with $h_{ij}$ the traceless ($\delta^{ij}h_{ij}=0$) and transverse ($\partial^i h_{ij}=0$) tensor perturbations obeying the condition, $|h_{ij}|\ll 1$.
In addition, $h_{+}$ and $h_{\times}$ denotes the two degrees of freedom of polarizations.  It is worth noting that the amount of tensor perturbations generated during an inflationary phase in the Einstein gravity was first and quantitatively correctly calculated by Starobinsky in a seminal paper \cite{Starobinsky:1979ty}. This is certainly the first observational test of inflation.

To quantize the tensor perturbations, we can Fourier transform $h_{ij}({\bf x},\eta)$ as
\begin{equation}
h_{ij}({\bf x},\eta) = \int \frac{d^3k}{(2\pi)^{3/2}} e^{i {\bf k}{\bf x}} \hat h_{ij}({\bf k},\eta).
\end{equation}
Additionally, we can also decompose $\hat h_{ij}({\bf k},\eta)$  in the Fourier space as
\begin{align}
\hat h_{ij}({\bf k},\eta) & =\hat h_{+}({\bf k},\eta) e^{+}_{ij}({\bf k})+\hat h_{\times} ({\bf k},\eta) e^{\times}_{ij}({\bf k}), \\
\hat h_{s}({\bf k},\eta) & = h_s(k,\eta)a_s({\bf k}) + h_s^\ast(k,\eta)a_s^\dag ({\bf k}).
\end{align}
Here $a_s({\bf k})$ and $a_s^\dag ({\bf k})$ are the creation and annihilation operators satisfying the commutation relation $[a_s({\bf k}),a_{s'}^\dag ({\bf k'})] = \delta_{ss'}\delta^3({\bf k}-{\bf k'})$. 
In addition, $e^{+}_{ij}({\bf k})$ and $e^{\times}_{ij}({\bf k})$ are symmetric polarization tensors obeying the following conditions $e^{s}_{ii}({\bf k})=0$ and $k_j e^{s}_{ij}({\bf k}) =0$ for all $s=+,\times$. 
For convenience, the normalization of $ e^{s}_{ij}({\bf k})$ is given by $e^{s}_{ij}({\bf k})e^{\ast s'}_{ij}({\bf k}) =\delta_{ss'}$. 
Note that the gravitational wave in  the KSW anisotropic inflation model lies on the $(x,y)$ plane due to  the rotational symmetry in the $(y,z)$ plane. 
Consequently, we can define the wave number vector in the Fourier space as ${\bf k}=k(\cos \theta,\sin \theta,0)$ with $\theta$ the angle between $\bf k$ and $x$-axis. Hence, it can be shown that \cite{Imprint}
\begin{eqnarray}
\hat h_{ij} = \frac{\hat h_+}{\sqrt{2}}\left( {\begin{array}{*{20}c}
   {\sin^2 \theta} & {-\sin \theta \cos \theta } & {0 }  \\
   {-\sin \theta \cos \theta } & {\cos^2 \theta } & {0}  \\
   {0 } & {0 } & {-1 }  \\
 \end{array} } \right)+ \frac{i \hat h_{\times}}{\sqrt{2}}\left( {\begin{array}{*{20}c}
   {0} & {0} & {-\sin \theta }  \\
   {0} & {0 } & {\cos \theta}  \\
   {-\sin \theta } & {\cos \theta } & {0 }  \\
 \end{array} } \right).
\end{eqnarray}
 In the de Sitter background space, the isotropic polarizations $h_{+}^{(0)}(k,\eta)$ and $h_{\times}^{(0)}(k,\eta)$ are both given by
\begin{equation}
h_{s}^{(0)} (k,\eta) =  \frac{\sqrt{2}H}{M_p k^{3/2}}\left(1+ik\eta\right)e^{-ik\eta},
\end{equation}
which are solutions of the following evolution equations for the gravitational wave amplitude,
\begin{equation}
h_{s}^{(0)''}+2\frac{a'}{a}h_{s}^{(0)'}+k^2h_{s}^{(0)}=0
\end{equation}
derived from the Einstein equations. Here the superscript $(0)$ denotes the de Sitter background. 
It is important to note that  the isotropic tensor power spectrum of the non-canonical scalar field model, $ {\cal P}^{h(0)}_{k,\text{nc}}$, can be shown to be identical to the result of canonical scalar field model since the gravity sector relative to the Ricci scalar in these two models is identical \cite{k-inflation,Baumann}. 
Indeed,  the isotropic power spectrum $ {\cal P}^{h(0)}_{k,\text{nc}}$ can be defined in terms of two-point correlation as
\begin{equation}
\langle 0 \vert \hat h_{ij}^{(0)}({\bf k}) \hat h_{ij}^{(0)}({\bf k'}) \vert 0\rangle = \frac{2\pi^2}{k^3}  \delta^3 ({\bf k}+{\bf k'}) {\cal P}^{h(0)}_{k,\text{nc}}({\bf k}),
\end{equation}
where 
\begin{equation} \label{def.of.Ph}
{\cal P}^{h(0)}_{k,\text{nc}}({\bf k})=  \left. \frac{2}{\pi^2} \frac{H^2}{M_p^2} \right|_{ k_\ast =a_\ast H_\ast}= 16 c_s \epsilon {\cal P}^{\zeta(0)}_{k,\text{nc}}({\bf k}).
\end{equation}
Next, we can calculate the correction due to the existence of the coupling term $h^2(\phi) F_{\mu\nu}F^{\mu\nu}$. 
As a result, all calculations of tensor perturbations shown in  Ref. \cite{Imprint} remain valid for the non-canonical scalar field (see Appendix \ref{app2} for details).
Indeed, we can show that
\begin{equation} \label{full-power-spectrum-tensor}
\langle 0 \vert \hat h_{ij}({\bf k}) \hat h_{ij}({\bf k'}) \vert 0\rangle =\frac{2\pi^2}{k^3}  \delta^3 ({\bf k}+{\bf k'})  \left({\cal P}^{h(0)}_{k,\text{nc}} +  \frac{4E_x^2 N_k^2}{\pi^2 M_p^4}\sin^2 \theta \right).
\end{equation}
 Hence, the full tensor power spectrum for the non-canonical scalar field  is given by
\begin{eqnarray} 
{\cal P}^h_{{k,\text{nc}}}  = && ~   {\cal P}^{h(0)}_{k,\text{nc}} +  \frac{4E_x^2 N_k^2}{\pi^2 M_p^4}\sin^2 \theta \nonumber\\
= && ~  {\cal P}^{h(0)}_{k,\text{nc}} \left(1- \frac{ \epsilon g_\ast^{0}}{4} \frac{N_k^2}{N_{c_sk}^2}\sin^2 \theta  \right)\nonumber\\
\simeq &&~  {\cal P}^{h(0)}_{k,\text{nc}} \left(1- \frac{ \epsilon g_\ast^{0}}{4} \sin^2 \theta  \right),
\end{eqnarray}
where we have used the approximation that $N_k \sim N_{c_s k}$ \cite{k-inflation}.

As a result, the corresponding tensor spectral index is given by 
\begin{equation}
n_t \equiv \left.\frac{d \ln  {\cal P}^h_{{k,\text{nc}}}}{d\ln k}\right|_{ k_\ast =a_\ast H_\ast} \simeq -2\epsilon,
\end{equation}
where  $\epsilon^2$ term is neglected for $\epsilon \ll 1$ during the inflationary phase \cite{Imprint}. 
\subsection{Full tensor-to-scalar ratio}
Finally, we are able to obtain the full tensor-to-scalar ratio \cite{Imprint,Lyth,Baumann} for the non-canonical scalar field coupled to the vector field model as
\begin{equation}   \label{general r}
r_{\rm nc} \equiv \frac{ {\cal P}^h_{k,\text{nc}}}{{\cal P}^\zeta_{k,\text{nc}} } = 16c_s \epsilon \frac{1- \frac{ 1}{4}\epsilon g_\ast^{0} \sin^2 \theta}{1-c_s^5 g_\ast^{0}  \sin^2 \theta}.
\end{equation}
Furthermore, taking the average value of  $\sin^2 \theta$ as $\langle \sin^2 \theta \rangle =2/3$ \cite{Imprint}, the  full tensor-to-scalar ratio reduces to
\begin{equation}   \label{general r-1}
r_{\rm nc} = 16c_s \epsilon \frac{6-\epsilon g_\ast^{0}}{6-4c_s^5 g_\ast^{0}}.
\end{equation}
It is clear that when $c_s=1$ then the tensor-to-scalar ratio for non-canonical scalar field will reduce to that for canonical scalar field, i.e., \cite{Imprint}
\begin{equation}
 r_{\rm nc} \to r = 16 \epsilon \frac{6-\epsilon g_\ast^0}{6-4g_\ast^0}.
\end{equation}

\section{CMB imprints of the non-canonical anisotropic inflation} \label{sec4}
The correlators of CMB observables take the following form  \cite{Imprint2,Imprint3}, 
\begin{eqnarray} \label{general correlations}
C^{X_1 X_2}_{l_1 l_2 m_1 m_2} \equiv && ~ \langle a_{l_1 m_1}^{X_1} a_{l_2 m_2}^{X_2 \ast} \rangle \nonumber\\
= &&~ 4\pi \int{\frac{dk}{k} \Delta_{l_1}^{i_1 X_1}(k) \Delta_{l_2}^{i_2 X_2}(k)} \int{ \left[_{i_1}Y^\ast_{l_1 m_1}(\theta,\phi) _{i_2}Y_{l_2 m_2}(\theta,\phi) \right] P^{i_1,i_2}(k,\theta,\phi) d \Omega}, 
\end{eqnarray}
where $\Delta$ is the transfer function, while $X$ represents the temperature anisotropy ($X_i =T$), the E-mode ($X_i=E$), or the B-mode ($X_i=B$). 
In addition, $_i Y_{lm} (\theta,\phi)$ is the spin-$i$-weighted spherical harmonics. 
Precise definitions will be listed in Appendix \ref{app4} for completeness. 
On the other hand, the power spectra of helicity bases, $P^{i_1,i_2}$, are given by \cite{Imprint,Imprint1,Imprint2}
\begin{eqnarray} 
P^{0,0}= && ~ {\cal P}_\zeta ,\\
P^{0,\pm 2}=&& ~ P^{\pm2,0}=\frac{1}{\sqrt{2}} P^{0,+} \equiv \frac{1}{\sqrt{2}}{\cal P}_{\zeta h_{+}}, \\
P^{\pm 2, \pm 2} = && ~\frac{1}{2} \left(P^{++}+P^{\times \times}\right) = \frac{1}{2}\left({\cal P}_{h_+} + {\cal P}_{h_\times}\right) \equiv {\cal P}^{\text{unp}}_h, \\
\label{gap}
P^{\pm 2, \mp 2} =&& ~ \frac{1}{2} \left(P^{++}-P^{\times \times}\right) = \frac{1}{2}\left({\cal P}_{h_+} - {\cal P}_{h_\times}\right)\equiv {\cal P}^{\text{pol}}_h ,
\end{eqnarray}
here we have noted that $ P^{0,\times} = P^{\times,0} \sim {\cal P}_{\zeta h_{\times}}=0$ due to the fact that $\epsilon^i_x \epsilon^j_z =0$. In addition, $P^{0,0}$,  $P^{0,\pm 2}$, $P^{\pm 2, \pm 2}$,  and $P^{\pm 2, \mp 2}$ correspond to the scalar perturbations, cross-correlations, tensor perturbations, and linear polarization, respectively. Note that  $P^{i_1,i_2} \sim \delta_{i_1 i_2}$ in the isotropic inflation, which implies that ${\cal P}_{\zeta h_{+}}$ and $ {\cal P}^{\text{pol}}_h $ must vanish identically when the spatial anisotropies do not show up. This should be a smoking gun for the anisotropic inflation.

Our next step is to define the explicit expression of $ {\cal P}_\zeta$, ${\cal P}^{\text{unp}}_h$, ${\cal P}^{\text{pol}}_h$, and ${\cal P}_{\zeta h_{+}}$ in the context of non-canonical extensions of the KSW model. As discussed earlier,  the full scalar and tensor spectra for non-canonical scalar fields are given by
\begin{eqnarray} \label{def of scalar power spectrum}
{\cal P}_\zeta= &&~{\cal P}^\zeta_{k,{\text{nc}}}    = {\cal P}^{\zeta(0)}_{k,{\text{nc}}} \left(1- c_s^5 g_\ast^0 \sin^2 \theta \right) ,\\ 
 \label{def of tensor power spectrum}
{\cal P}^{\text{unp}}_h = && ~  {\cal P}^h_{{k,\text{nc}}} =  {\cal P}^{h(0)}_{k,\text{nc}} \left(1- \frac{ \epsilon g_\ast^0}{4} \sin^2 \theta \right).
\end{eqnarray}
Furthermore, we are able define the cross-correlation between curvature perturbations and the '$+$' mode of gravitational waves, i.e., ${\cal P}_{\zeta h_{+}}$, to be \cite{Imprint,Imprint1,Imprint2,Imprint3} (see the Appendix \ref{app3} for specific details)
\begin{equation} \label{P-cross}
{\cal P}_{\zeta h_{+}} =  -c_s^2 \frac{2 E_x^2 N_{c_sk} N_k }{\sqrt{2} \pi^2 \epsilon M_p^4}  \sin^2 \theta  =c_s^3 \sqrt{2} \epsilon g_\ast^0     {\cal P}^{\zeta(0)}_{k,{\text{nc}}} \sin^2 \theta.
\end{equation}
It is noted that ${\cal P}_{h_+} = {\cal P}_{h_\times}$ up to the second order term (see the Appendix \ref{app2} for the proof), which results that ${\cal P}^{\text{pol}}_h \simeq 0$ according to the definition in Eq. (\ref{gap}). Therefore, we need to consider higher order terms of ${\cal P}_{h_+}$ and ${\cal P}_{h_\times}$ to see if there is any difference between them. And the gap, if exists, will be exactly the value of  ${\cal P}^{\text{pol}}_h$ as shown in Eq. (\ref{gap}).  
As a result, we can define  ${\cal P}^{\text{pol}}_h$ up to the fourth order term as \cite{Imprint,Imprint1,Imprint2,Imprint3} (see Appendix \ref{app3} for details)
\begin{equation} \label{def.of.P-linear}
{\cal P}^{\text{pol}}_h =  g_l  ~   {\cal P}^{h(0)}_{k,\text{nc}} \sin^4 \theta ,
\end{equation}
where 
\begin{equation} \label{def.gl}
g_l \sim  {\cal O} \left({\epsilon^2 (g_\ast^0)^2}\right) .
\end{equation}
 For each $ {\cal P}_\zeta$, ${\cal P}^{\text{unp}}_h$, ${\cal P}^{\text{pol}}_h$, or ${\cal P}_{\zeta h_{+}}$, we obtain the corresponding correlations which could be observed. 
Moreover, information from observed correlations might help us to reveal the effect of the statistical anisotropy. 
In particular, the anisotropy in the scalar or tensor perturbations might be figured out in the correlations associated with  $ {\cal P}_\zeta$ or ${\cal P}^{\text{unp}}_h$, respectively. 
In addition, the anisotropy in the  linear polarization or  cross-correlation might be seen in the correlations corresponding to ${\cal P}^{\text{pol}}_h$ or ${\cal P}_{\zeta h_{+}}$, respectively \cite{Imprint,Imprint1,Imprint2,Imprint3}.
\subsection{Anisotropy in the scalar perturbations}
According to the definition in Eq. (\ref{general correlations}), we have the corresponding correlations $TT$, $TE$, and $EE$  in the scalar perturbations ($i_1 =i_2 =0$):
\begin{equation} \label{general correlations-scalar}
C^{X_1 X_2}_{l_1 l_2 m_1 m_2}=  4\pi \int{\frac{dk}{k} \Delta_{l_1}^{0 X_1}(k) \Delta_{l_2}^{0 X_2}(k)} \int{ \left[_{0}Y^\ast_{l_1 m_1}(\theta,\phi) _{0}Y_{l_2 m_2}(\theta,\phi) \right] P^{0,0}({ k,\theta,\phi}) d \Omega}. 
\end{equation}
Next, we expand the anisotropic scalar power spectrum $P^{0,0}$ as the spherical harmonics \cite{Imprint2,Imprint3}
\begin{equation}
P^{0,0} \equiv {\cal P}^\zeta_{k,{\text{nc}}}  = \sum\limits_{L,M} a^{00}_{LM}(k)~ {_0 Y_{LM}},
\end{equation}
with $ a^{00}_{LM}(k) =0$ for odd $L$. 
Furthermore, using the relation \cite{Imprint2,Imprint3},
\begin{equation}
\int {  _0Y_{LM} { _{-s}Y^{\ast}_{l_1m_1}} {_{-s} Y_{l_2 m_2}}~d \Omega} = \sqrt{\frac{(2L+1)(2l_2+1)}{4\pi (2l_1+1)}} {\cal C}^{l_1 m_1}_{LM l_2 m_2} {\cal C}^{l_1 s}_{L0l_2 s},
\end{equation}
we can simplify the Eq. (\ref{general correlations-scalar}) as
\begin{eqnarray} \label{simplify correlations-scalar}
C^{X_1 X_2}_{l_1 l_2 m_1 m_2}=  4\pi \int{\frac{dk}{k} \Delta_{l_1}^{0 X_1}(k) \Delta_{l_2}^{0 X_2}(k)   \sum\limits_{L,M}a^{00}_{LM}(k) \sqrt{\frac{(2L+1)(2l_2+1)}{4\pi (2l_1+1)}}  {\cal C}^{l_1 m_1}_{LM l_2 m_2} {\cal C}^{l_1 0}_{L0l_2 0}   },\nonumber\\
\end{eqnarray}
where ${\cal C}$ stands for the Clebsch-Gordan coefficient defined as
\begin{equation}
{\cal C}^{l_1 m_1}_{LMl_2 m_2}= \langle L, l_2; M, m_2| L, l_2; l_1, m_1 \rangle.
\end{equation}
Noting the definition of $_0Y_{20}(\theta,\phi)$:
\begin{equation} \label{0Y20}
_0 Y_{20}(\theta,\phi) = \frac{1}{4} \sqrt{\frac{5}{\pi}} \left(3 \cos^2 \theta -1 \right) = \frac{3}{4}\sqrt{\frac{5}{\pi}} \left(\frac{2}{3}-\sin^2 \theta \right),
\end{equation}
and the definition of ${\cal P}^\zeta_{k,{\text{nc}}}  $ shown in Eq. (\ref{def of scalar power spectrum}),
we can define the component $a^{00}_{20}$, which characterizes the statistical anisotropy, to be
\begin{equation}
a^{00}_{20}(k) = \frac{4}{3}\sqrt{\frac{\pi}{5}} c_s^5 g_\ast^0  {\cal P}^{\zeta(0)}_{k,{\text{nc}}}(k).
\end{equation}
As a result, Eq. (\ref{simplify correlations-scalar}) implies the anisotropic contribution to the scalar perturbation due to the non-vanishing $a^{00}_{20}(k)$: 
\begin{eqnarray} \label{off-diagonal-scalar}
C^{X_1 X_2}_{l_1 l_2 m_1 m_2}=\frac{  8\pi  c_s^5 g_\ast^0}{3}\sqrt{\frac{2l_2+1}{2l_1+1}} {\cal C}^{l_1 m_1}_{20 l_2 m_2} {\cal C}^{l_1 0}_{20l_2 0}} \int{\frac{dk}{k} \Delta_{l_1}^{0 X_1}(k) \Delta_{l_2}^{0 X_2}(k)   {\cal P}^{\zeta(0)}_{k,{\text{nc}}}(k),
\end{eqnarray}
while the isotropic contribution is mainly due to $a^{00}_{00}(k)$ associated with $_0Y_{00}=1/(2\sqrt{\pi})$. 
 In addition, these anisotropic contribution  are non-zero only for even $(l_1 -l_2)$, e.g., $l_2 =l_1$ or $l_2 =l_1\pm 2$. 
Note again that in the scalar perturbations $X_{1,2}$ can only be $T$ or $E$ and the definition of isotropic scalar power spectrum ${\cal P}^{\zeta(0)}_{k,{\text{nc}}}$ has been shown in Eq. (\ref{def.of.Pzeta}).
\subsection{Anisotropy in the tensor perturbations}
Now, we would like to define the correlations in the tensor perturbations:
\begin{equation} \label{general correlations-tensor}
C^{X_1 X_2 \pm}_{l_1 l_2 m_1 m_2}=   4\pi \int{\frac{dk}{k} \Delta_{l_1}^{2 X_1}(k) \Delta_{l_2}^{2 X_2}(k)}  \int{ \left[_{-2}Y^\ast_{l_1 m_1}(\theta,\phi) _{-2}Y_{l_2 m_2}(\theta,\phi) ~\pm ~_{2}Y^\ast_{l_1 m_1}(\theta,\phi) _{2}Y_{l_2 m_2}(\theta,\phi)\right]  {\cal P}^{\text{unp}}_h (k,\theta,\phi) d \Omega} ,
\end{equation} 
where '$+$'  sign corresponds to $TT$, $EE$, $TE$, and $BB$ correlations; while '$-$'  sign corresponds to $TB$ and $EB$ correlations \cite{Imprint2,Imprint3}. Similar to the method used in the scalar perturbations, we will expand the spectrum ${\cal P}^{\text{unp}}_h $ into spherical harmonics given by
\begin{equation}
{\cal P}^{\text{unp}}_h  =  \sum\limits_{L,M} a^{\text{unp}}_{LM}(k)~ {_0 Y_{LM}},
\end{equation}
with $a^{\text{unp}}_{LM}(k) =0$ for odd $L$ \cite{Imprint}. Hence,
\begin{equation} 
 C^{X_1 X_2 \pm}_{l_1 l_2 m_1 m_2}=  4\pi \int{\frac{dk}{k} \Delta_{l_1}^{2 X_1}(k) \Delta_{l_2}^{2 X_2}(k)}  \sum\limits_{L,M} a^{\text{unp}}_{LM}(k) \sqrt{\frac{(2L+1)(2l_2+1)}{4\pi (2l_1+1)}} {\cal C}^{l_1 m_1}_{LM l_2 m_2}  \left[ {\cal C}^{l_1 2}_{L0 l_2 2} \pm {\cal C}^{l_1(-2)}_{L0l_2(-2)}\right].
\end{equation} 
Using the definition of ${\cal P}^{\text{unp}}_h $ shown in Eq. (\ref{def of tensor power spectrum}) and that of $_0Y_{20}(\theta,\phi)$ as shown in Eq. (\ref{0Y20}) in the scalar perturbations, we can obtain
\begin{equation}
 a^{\text{unp}}_{20}(k) = \frac{4}{3}\sqrt{\frac{\pi}{5}} \frac{\epsilon g_\ast^0}{4}  {\cal P}^{h(0)}_{k,\text{nc}}(k).
\end{equation}
It turns out that while the isotropic contribution is mainly due to $a^{\text{unp}}_{00}(k)$ associated with $_0Y_{00}=1/(2\sqrt{\pi})$, the anisotropic contribution to the tensor perturbation due to the non-vanishing $a^{\text{unp}}_{20}(k) $ is given by
\begin{equation} \label{off-diagonal-tensor}
 C^{X_1 X_2 \pm}_{l_1 l_2 m_1 m_2}=  \frac{2\pi \epsilon g_\ast^0}{3} \sqrt{\frac{2l_2+1}{2l_1+1}}  {\cal C}^{l_1 m_1}_{20 l_2 m_2}  \left[ {\cal C}^{l_1 2}_{20 l_2 2} \pm {\cal C}^{l_1(-2)}_{20l_2(-2)}\right]  \int{\frac{dk}{k} \Delta_{l_1}^{2 X_1}(k) \Delta_{l_2}^{2 X_2}(k)  {\cal P}^{h(0)}_{k,\text{nc}}(k)},
\end{equation} 
where ${\cal P}^{h(0)}_{k,\text{nc}}(k)$ has been defined in Eq. (\ref{def.of.Ph}).
As a result, the correlations $C^{X_1 X_2 +}_{l_1 l_2 m_1 m_2}$ ($TT$, $EE$, $TE$, and $BB$) are non-vanishing only for even $(l_1-l_2)$, e.g., $l_2 =l_1$ or $l_2 =l_1 \pm 2$; while the correlations $C^{X_1 X_2 -}_{l_1 l_2 m_1 m_2}$ ($TB$ and $EB$) are non-vanishing only for odd $(l_1-l_2)$, e.g., $l_2 =l_1 \pm 1$. 
\subsection{Anisotropy in the cross-correlations}
As a result, the following $TT$, $EE$, and $TE$ correlations induced by the cross-correlations  are given by \cite{Imprint2,Imprint3}
\begin{eqnarray} \label{correlations-cross-1}
 C^{X_1 X_2 }_{l_1 l_2 m_1 m_2} 
 = &&~ 8 \pi\epsilon  c_s^3 g_\ast^0     \int{\frac{dk}{k}} {\cal P}^{\zeta(0)}_{k,{\text{nc}}}({ k})  \left\{  \Delta^{0 X_1}_{l_1}(k) \Delta^{2X_2}_{l_2}(k) \left(\alpha^{-2}_{l_1 +2, m_1} \delta_{l_2,l_1+2} + \alpha^0_{l_1,m_1} \delta_{l_2,l_1} \right. \right. \nonumber\\
&& \left. \left. +\alpha^{+2}_{l_1-2, m_1} \delta_{l_2,l_1 -2} \right) \delta_{m_1 m_2}   + (-1)^{l_1 +l_2 +m_1 +m_2} \Delta^{2X_1}_{l_1}(k) \Delta^{0 X_2}_{l_2}(k) \left(\alpha^{-2}_{l_2 +2, -m_2} \delta_{l_1,l_2+2} \right. \right. \nonumber\\
&& \left.\left.    + \alpha^0_{l_2,-m_2} \delta_{l_1,l_2}+\alpha^{+2}_{l_2-2, -m_2} \delta_{l_1,l_2 -2} \right)  \delta_{-m_2, -m_1}  \right\}
\end{eqnarray} 
along with the $TB$ and $EB$ spectra defined as \cite{Imprint2,Imprint3}
\begin{eqnarray}\label{correlations-cross-2}
 C^{X_1 B }_{l_1 l_2 m_1 m_2} = &&~ \frac{4 \pi}{\sqrt{2}}  \int{\frac{dk}{k}} \Delta^{0 X_1}_{l_1}(k) \Delta^{2B}_{l_2} (k)\int Y^\ast_{l_1 m_1} \left[ _{-2}Y_{l_2 m_2}-_{+2}Y_{l_2 m_2}  \right] {\cal P}_{\zeta h_{+}}d \Omega \nonumber\\
=&&~   8 \pi \epsilon c_s^3 g_\ast^0    \int{\frac{dk}{k}} {\cal P}^{\zeta(0)}_{k,{\text{nc}}} ({ k}) \Delta^{0 X_1}_{l_1}(k) \Delta^{2B}_{l_2} (k)\left(\beta^{-1}_{l_1+1,m_1} \delta_{l_2,l_1+1} +\beta^{+1}_{l_1-1,m_1}\delta_{l_2,l_1-1}\right) \delta_{m_1 m_2}, 
\end{eqnarray}
where we have used the definition of ${\cal P}^{\zeta(0)}_{k,{\text{nc}}}$  and $ {\cal P}_{\zeta h_{+}}$  shown in Eqs. (\ref{def.of.Pzeta}) and (\ref{P-cross}).  In addition, the $\alpha$ and $\beta$  functions have been defined  \cite{Imprint2} as
\begin{eqnarray} \label{alpha+2}
\alpha^{+2}_{l,m} \equiv && ~ \sqrt{\frac{l(l-1)(l+m+1)(l-m+1)(l+m+2)(l-m+2)}{(l+1)(l+2)(2l+1)(2l+3)^2(2l+5)}}, \\
 \label{alpha-2}
\alpha^{-2}_{l,m} \equiv&& ~ \sqrt{\frac{(l+1)(l+2)(l+m)(l-m)(l+m-1)(l-m-1)}{(l-1)l(2l-3)(2l-1)^2(2l+1)}}, \\
 \label{alpha0}
\alpha^0_{l,m}\equiv&&~ \frac{2\left[3m^2 -l(l+1)\right]}{(2l-1)(2l+3)}\sqrt{\frac{(l-1)(l+2)}{l(l+1)}}, \\
 \label{beta+1}
\beta^{+1}_{l,m}\equiv && ~ 2m \sqrt{\frac{(l-1)(l+m+1)(l-m+1)}{l(l+1)(l+2)(2l+1)(2l+3)}}, \\
 \label{beta-1}
\beta^{-1}_{l,m}\equiv&&~ -2m \sqrt{\frac{(l+2)(l+m)(l-m)}{(l-1)l(l+1)(2l+1)(2l-1)}}.
\end{eqnarray}
\subsection{Anisotropy in the linear polarization}
We will focus on the correlations induced by the linear polarization in this subsection. 
As a result, the following $TT$, $EE$, $BB$, and $TE$ spectra are given by  \cite{Imprint2,Imprint3}
\begin{eqnarray} \label{correlations-linear-1}
C^{X_1 X_2}_{l_1 l_2 m_1 m_2} = &&~ 4 \pi  \int{\frac{dk}{k}} \Delta^{2 X_1}_{l_1}(k) \Delta^{2 X_2}_{l_2} (k) \int \left[{ _{-2}Y^\ast_{l_1 m_1}} {_{+2}Y_{l_2 m_2} } + {_{+2}Y^\ast_{l_1 m_1}} {_{-2}Y_{l_2 m_2} } \right]{\cal P}^{\text{pol}}_h  d \Omega \nonumber\\
= &&~8 \pi g_l  \int{\frac{dk}{k}}   {\cal P}^{h(0)}_{k,\text{nc}}(k) \Delta^{2 X_1}_{l_1}(k) \Delta^{2 X_2}_{l_2} (k)  \left\{ \alpha^{+2}_{l_1, m_1} \alpha^{-2}_{l_1+4,m_1} \delta_{l_2, l_1+4} \right. \nonumber\\
&& \left. + \left(\alpha^{+2}_{l_1, m_1} \alpha^{0}_{l_1+2,m_1} +\alpha^0_{l_1,m_1} \alpha^{-2}_{l_1+2,m_1}-\beta^{+1}_{l_1, m_1} \beta^{-1}_{l_1+2,m_1} \right) \delta_{l_2,l_1+2} \right.\nonumber\\
&& \left. +\left[(\alpha^{+2}_{l_1, m_1})^2 + (\alpha^0_{l_1, m_1})^2 + (\alpha^{-2}_{l_1 ,m_1})^2- (\beta^{-1}_{l_1, m_1})^2 - (\beta^{+1}_{l_1, m_1})^2\right] \delta_{l_2 , l_1} \right. \nonumber\\
&& \left. +\left(\alpha^0_{l_1,m_1}\alpha^{+2}_{l_1-2,m_1}+ \alpha^{-2}_{l_1,m_1}\alpha^0_{l_1-2,m_1} - \beta^{-1}_{l_1,m_1}\beta^{+1}_{l_1-2,m_1} \right)\delta_{l_2,l_1-2} \right. \nonumber\\
&& \left. +\alpha^{-2}_{l_1,m_1} \alpha^{+2}_{l_1-4,m_1} \delta_{l_2,l_1-4}\right\} \delta_{m_1,m_2},
\end{eqnarray}
here we have used the definition of ${\cal P}^{\text{pol}}_h $ shown in Eq. (\ref{def.of.P-linear}) and that of $\alpha$ and $\beta$ functions defined in Eqs. (\ref{alpha+2}), (\ref{alpha-2}), (\ref{alpha0}), (\ref{beta+1}), and (\ref{beta-1}). In addition, the isotropic  tensor power spectra ${\cal P}^{h(0)}_{k,\text{nc}}$ has been defined in Eq. (\ref{def.of.Ph}).

On the other hand, the following  $TB$ and $EB$ spectra are defined as  ~\cite{Imprint2,Imprint3}
\begin{eqnarray}\label{correlations-linear-2}
C^{X_1 X_2}_{l_1 l_2 m_1 m_2} = &&~ 4 \pi  \int{\frac{dk}{k}} \Delta^{2 X_1}_{l_1}(k) \Delta^{2 X_2}_{l_2} (k) \int \left[{ _{-2}Y^\ast_{l_1 m_1}} {_{+2}Y_{l_2 m_2} } - {_{+2}Y^\ast_{l_1 m_1}} {_{-2}Y_{l_2 m_2} } \right]{\cal P}^{\text{pol}}_h  d \Omega \nonumber\\
= &&~8 \pi g_l  \int{\frac{dk}{k}}   {\cal P}^{h(0)}_{k,\text{nc}}(k) \Delta^{2 X_1}_{l_1}(k) \Delta^{2 X_2}_{l_2} (k) \left[ \left(\beta^{+1}_{l_1,m_1}\alpha^{-2}_{l_1+3,m_1} - \alpha^{+2}_{l_1,m_1} \beta^{-1}_{l_1+3,m_1 }\right)\delta_{l_2,l_1+3} \right.\nonumber\\
&& \left. +\left(\beta^{+1}_{l_1,m_1} \alpha^0_{l_1+1,m_1} + \beta^{-1}_{l_1,m_1} \alpha^{-2}_{l_1+1,m_1} -\alpha^{+2}_{l_1,m_1}\beta^{+1}_{l_1+1,m_1} -\alpha^0_{l_1,m_1} \beta^{-1}_{l_1+1,m_1}\right)\delta_{l_2,l_1+1} \right.\nonumber\\
&& \left. + \left(\beta^{+1}_{l_1,m_1} \alpha^{+2}_{l_1-1,m_1}+\beta^{-1}_{l_1,m_1}\alpha^0_{l_1-1,m_1} -\alpha^{0}_{l_1,m_1} \beta^{+1}_{l_1-1,m_1} -\alpha^{-2}_{l_1,m_1} \beta^{-1}_{l_1-1,m_1}\right)\delta_{l_2,l_1-1} \right. \nonumber\\
&& \left. + \left(\beta^{-1}_{l_1,m_1}\alpha^{+2}_{l_1-3,m_1} -\alpha^{-2}_{l_1,m_1}\beta^{+1}_{l_1-3,m_1}\right)\delta_{l_2,l_1-3}\right] \delta_{m_1,m_2}.
\end{eqnarray}
Note that the coupling $g_l$ will be chosen as shown in Eq. (\ref{def.gl}), i.e., $g_l \simeq \epsilon^2 (g_\ast^0)^2/16$.
\subsection{Numerical results}
Note that the transfer function $\Delta$ can be calculated numerically by  performing the standard Boltzmann codes  \cite{Imprint2,Imprint3}. 
In addition, for the transfer function $\Delta=\Delta(k)$, we can also apply the well-known codes built for isotropic inflation to the anisotropic inflationary models. 
In particular, the {\it Cosmic Linear Anisotropy Solving System} (CLASS)  package \cite{code} will be employed to solve the transfer functions and plot the $TT$, $EE$,  $BB$,  $TE$, $TB$, and $EB$  correlations \cite{Imprint3}. 

Note that $c_s^2 \simeq 1$ for the SDBI inflationary model since $\gamma \simeq 1$ is required to endorse inflationary solutions \cite{Do:2016ofi}.  
Hence, all numerical results of the SDBI model is will be closely equal to the result of the canonical KSW model \cite{Imprint,Imprint2}. 
However, $c_s^2$ is arbitrary for  the DBI inflationary model free from any constraint on $\gamma$ \cite{WFK1}. 
Hence, significant gaps could show up for the $TT$, $EE$,  $BB$,  $TE$, $TB$, and $EB$  correlations between the DBI model and the canonical KSW model.

Note also that $g_\ast$ and $g_\ast^0$ are always negative as shown in Eqs. (\ref{definition_of_g}) and (\ref{definition_of_g0}). 
In addition, the latest joint analysis on primordial gravitational waves of BICEP2 and Keck array using the Planck, WMAP, and new BICEP2/Keck observations through the 2015 season provides an upper bound for the tensor-to-scalar ratio as $r_{\rm 0.05} \leq 0.07$ at $95\%$ confidence level (CL) \cite{Ade:2018gkx}. Furthermore, Planck collaboration has announced the  most recent data at the pivot scale $k_\ast =0.05\text{Mpc}^{-1}$, e.g., $n_s = 0.9649 \pm0.0042$ at $68 \%$ CL and $10^9 A_s = 2.100 \pm 0.030$ at $68 \%$ CL \cite{Planck2}. Hence, we will use the Planck 2018 data \cite{Planck2} for our numerical results.

Note that $r$ and $g_\ast^0$ were chosen in Ref. \cite{Imprint2} as $0.3$, which seem to be inconsistent with the recent observational data. 
Hence, we will choose a more reliable set of parameters with $g_\ast^0=-0.03$, according to Refs. \cite{data,data1,data2,data3,data4}, and $r_{\rm nc}=0.03$, according to Ref. \cite{Ade:2018gkx}. 
We will also set the coupling $g_l$ as $g_l \simeq \epsilon^2 (g_\ast^0)^2/16$ in order to define the following spectra in linear polarization. 
According to the relation shown in Eq. (\ref{general r-1}), we have $\epsilon \simeq 0.0019$ for $c_s=1$, while $\epsilon \simeq 0.0187$ for $c_s=0.1$. 

\begin{figure}[hbtp]
\begin{center}
{\includegraphics[height=50mm]{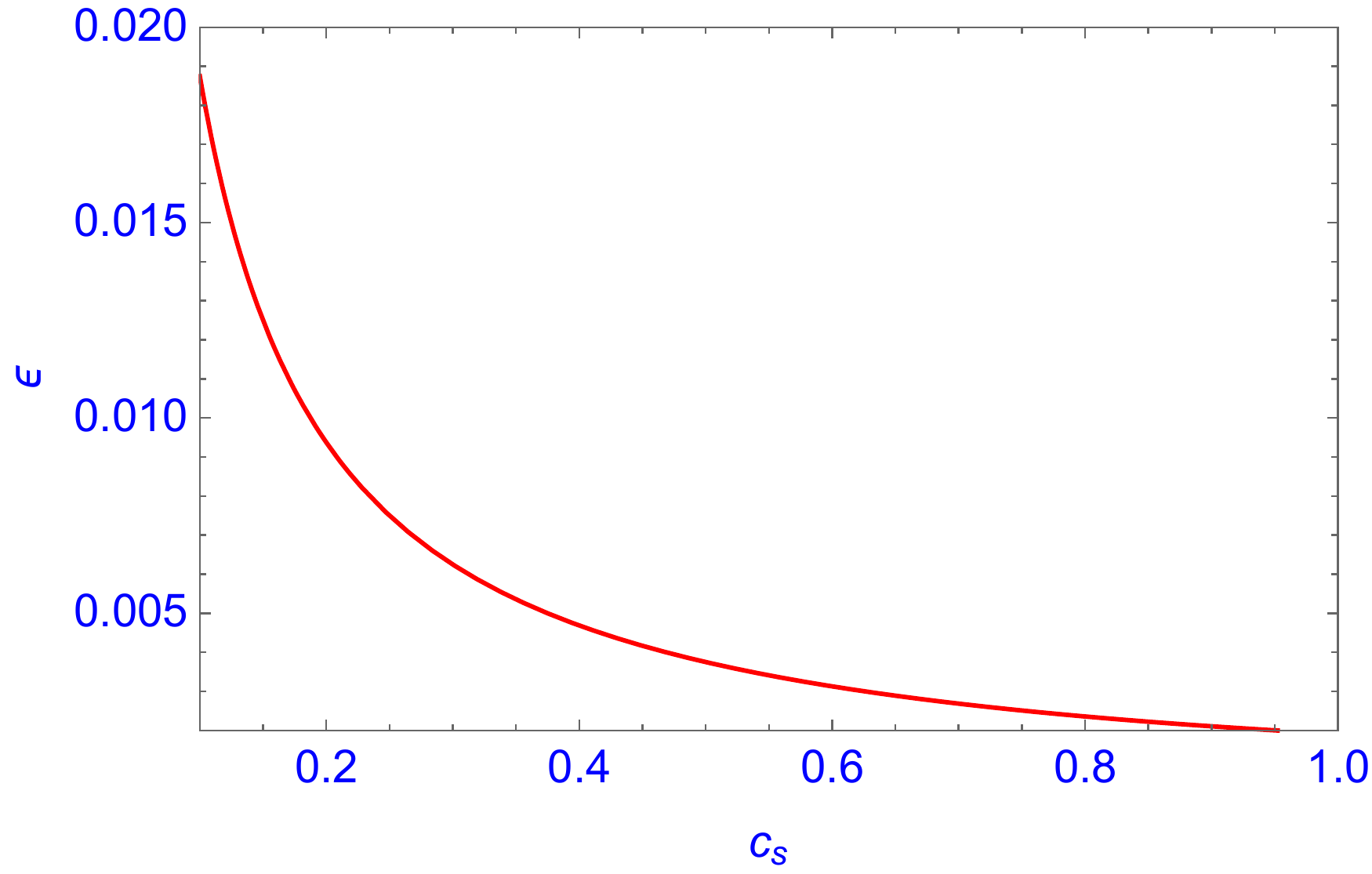}}\\
\caption{$\epsilon$ as a function of $c_s$. }
\label{fig0}
\end{center}
\end{figure}

Eq. (\ref{general r-1}) implies that $c_s$ and $\epsilon$ are not proportional to each other. 
Indeed, we can plot $\epsilon$ as a function of $c_s$ in Fig. \ref{fig0}.

In addition, we will plot the most significant (high amplitude) spectra according to the definitions given by Eqs. (\ref{off-diagonal-scalar}),  (\ref{off-diagonal-tensor}), (\ref{correlations-cross-1}), (\ref{correlations-cross-2}), (\ref{correlations-linear-1}), and (\ref{correlations-linear-2}). 
All figures are plotted in logarithm scales, i.e. $\ln |y|$ vs $\ln l$ for all multipole moment $l \le 1000$ with $y$ the magnitude of various spectra. 
In addition,  dashed, dotted, or dotted-dashed curves will be used to signify the region where the spectra is negative, i.e. $ y <0$. 

\begin{figure}[hbtp]
\begin{center}
{\includegraphics[height=50mm]{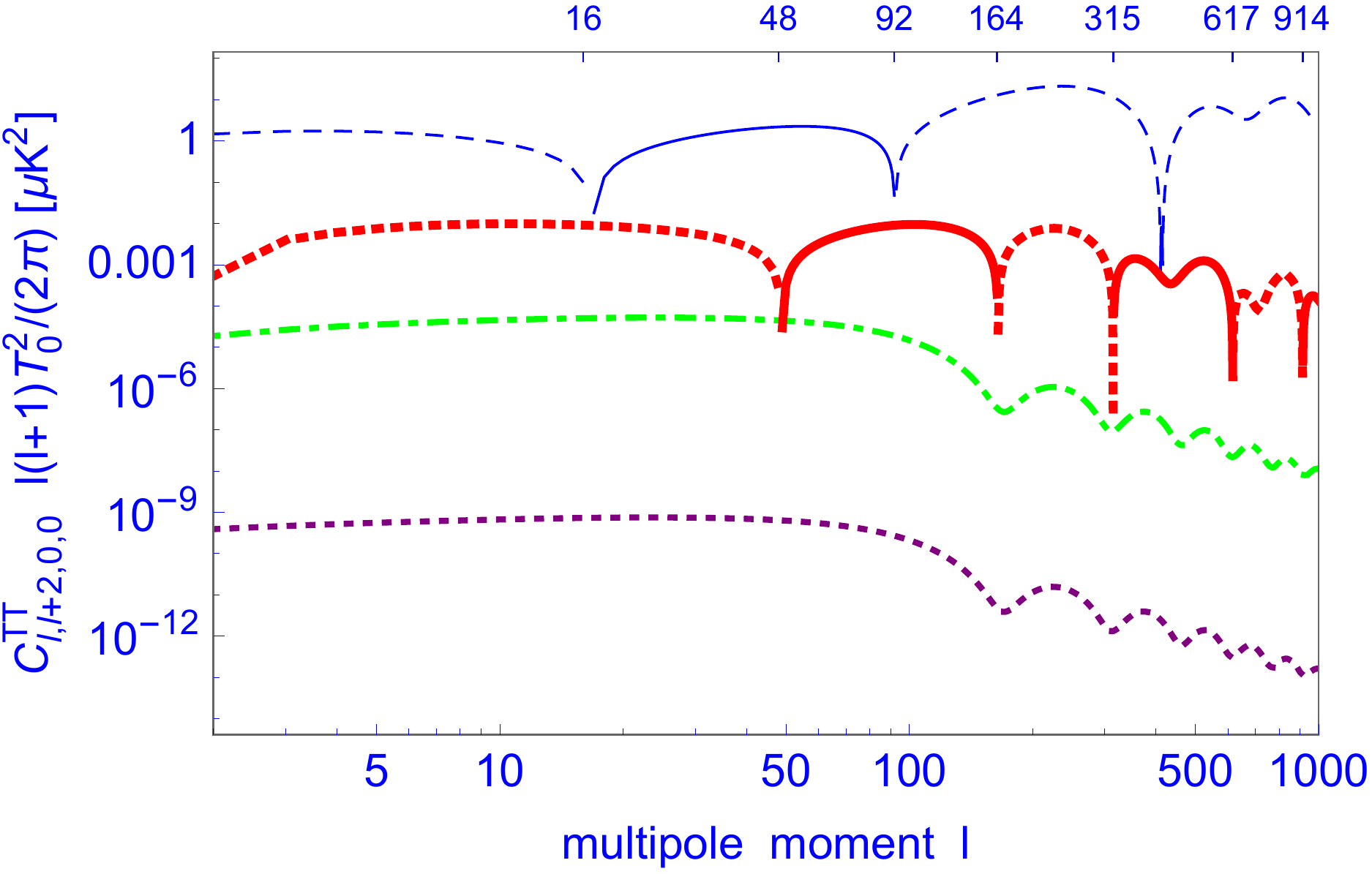}}\quad
{\includegraphics[height=50mm]{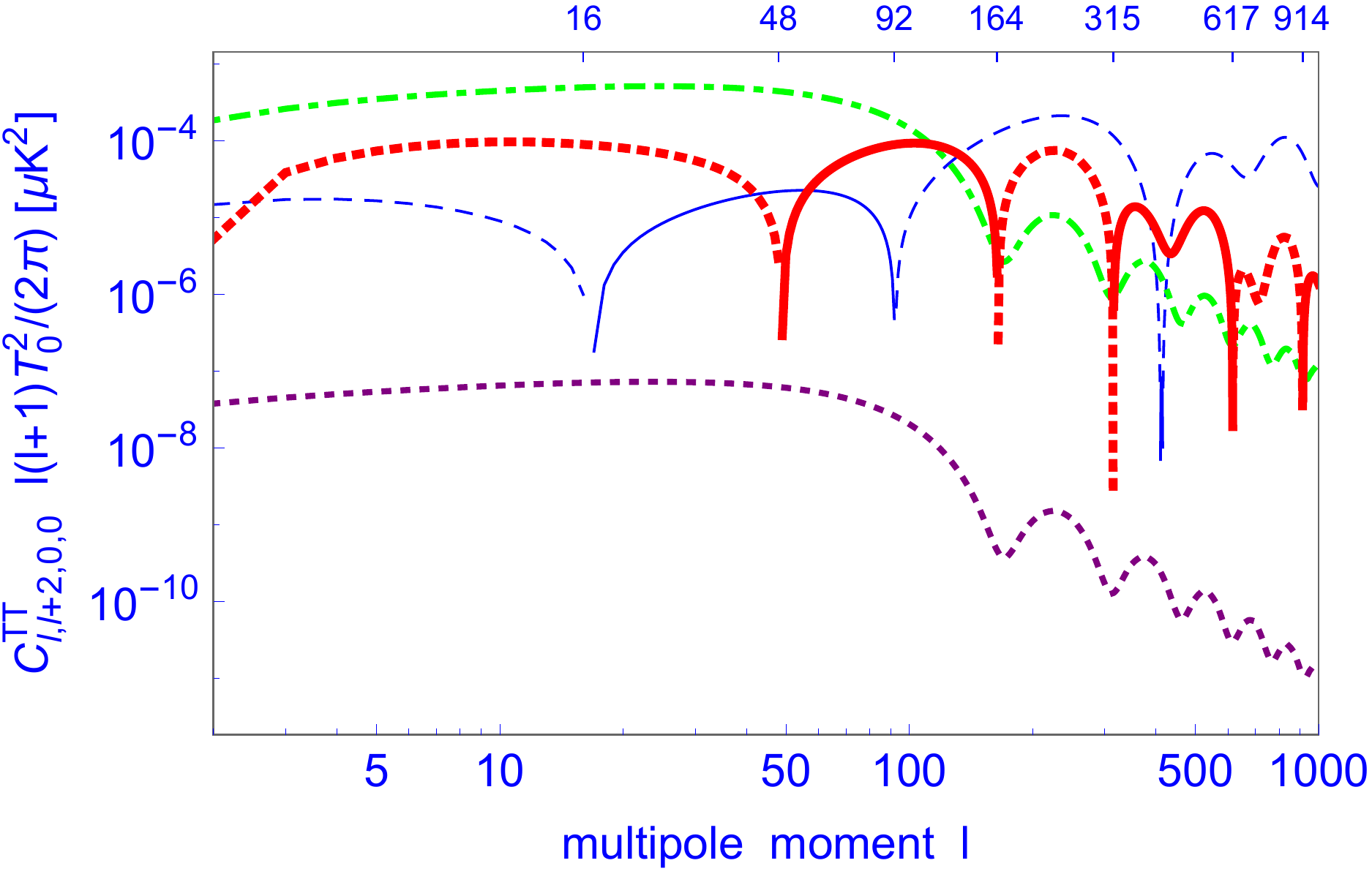}}\\
\caption{The magnitude of the $TT$ spectra $C^{TT}_{l,l+2,0,0}$ induced by the anisotropy in the scalar perturbations (the upper blue dashed-solid curve); by the  anisotropy in the tensor perturbations (the green dotted-dashed curve);  by the cross-correlations (the thicker red dotted-solid curve); and by the linear polarization  (the bottom purple dotted curve) are shown respectively. 
The dotted (red and purple), dashed (blue), and dotted-dashed (green) curves denotes the regions where $C^{TT}_{l,l+2,0,0}<0$.  
The left hand side and right hand side figures correspond to the canonical model with $c_s=1$ and the non-canonical model with $c_s=0.1$, respectively. 
The left figure implies that the $TT$ spectrum associated with the scalar perturbations dominates over the others for the canonical scalar field. 
However, this will not be true for the non-canonical scalar field as shown in the right figure. In addition, the plots display the sign-change points of $C^{TT}_{l,l+2,0,0}$ induced by the anisotropy in the scalar perturbations (at $l=16,~92$) and $C^{TT}_{l,l+2,0,0}$ induced by the cross-correlations (at $l=48,~164,~315,~617,~914$).}
\label{fig1}
\end{center}
\end{figure}

According to these numerical plots, the magnitude of all spectra for both $c_s=1$ and $c_s <1$ is much lower than the result in Ref. \cite{Imprint2}.
This is because that the $r$ and $g_\ast^0$ chosen in this paper are all of ${\cal O}(10^{-2})$, while $r$ and $g_\ast^0$ in Ref. \cite{Imprint2} are all of ${\cal O}(10^{-1})$. 
As shown in Fig. \ref{fig1},  two of four $TT$ spectra decrease accordingly with different rates when $c_s$ decreases from $1$ to $0.1$.
One of them is induced by the anisotropy in the scalar perturbations (blue dashed-solid curve).
The other one is induced by the cross-correlations (thick red dotted-solid curve). 
The spectra induced by the anisotropy of the scalar perturbations decreases with fastest speed since it is proportional to $c_s^5$ as shown in Eq. (\ref{off-diagonal-scalar}). 
On the other hand, the remaining $TT$ spectra induced by the tensor perturbations (green dotted-dashed curve) and by linear polarization (thin purple dotted curve) increase, however, when $c_s$ decreases. 
This is because that they depend on $\epsilon$, which will increase from $0.0019$ to  $0.0187$ when $c_s$ decreases from $1$ to $0.1$ as shown above. 

\begin{figure}[hbtp]
\begin{center}
{\includegraphics[height=50mm]{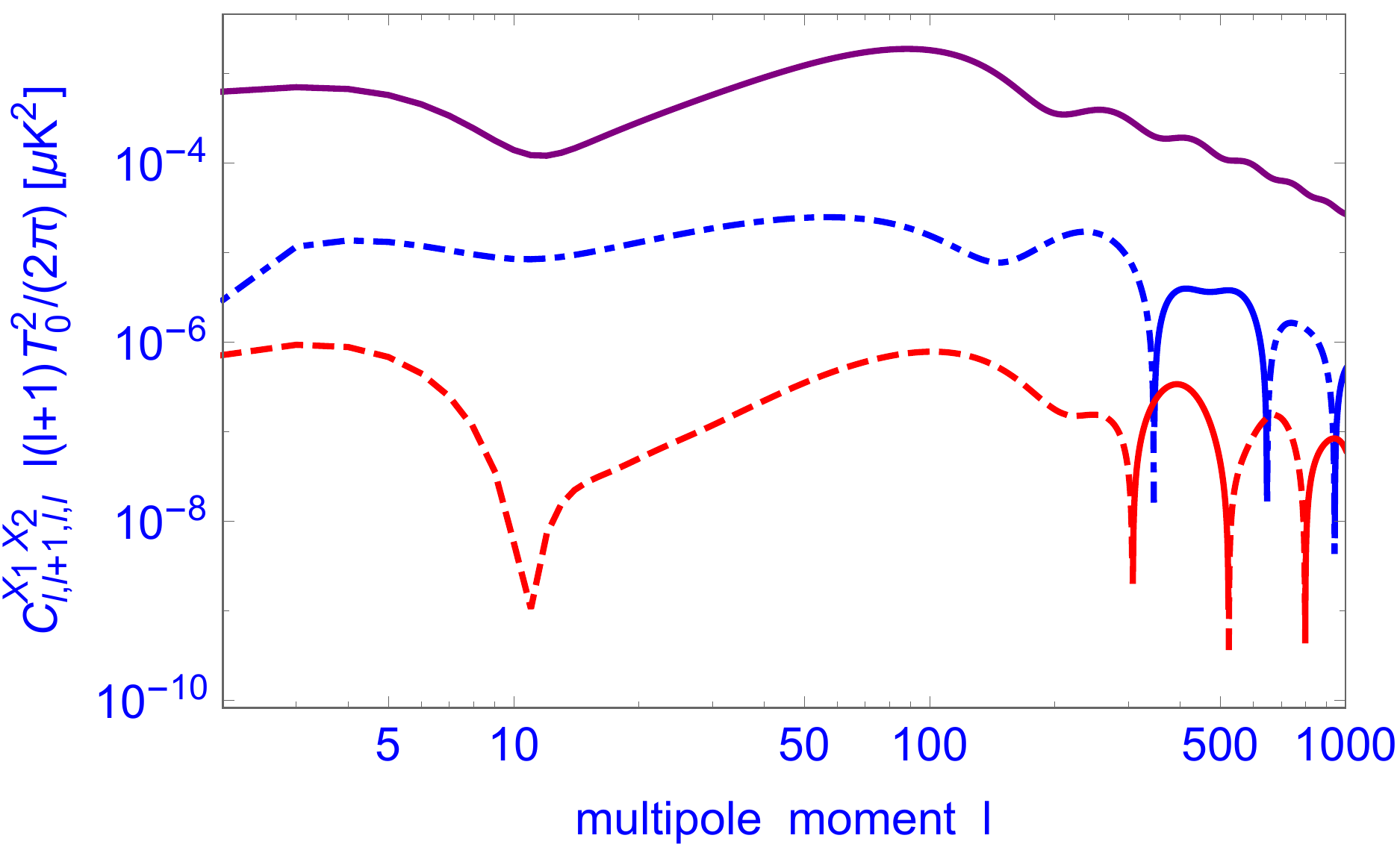}}\quad
{\includegraphics[height=50mm]{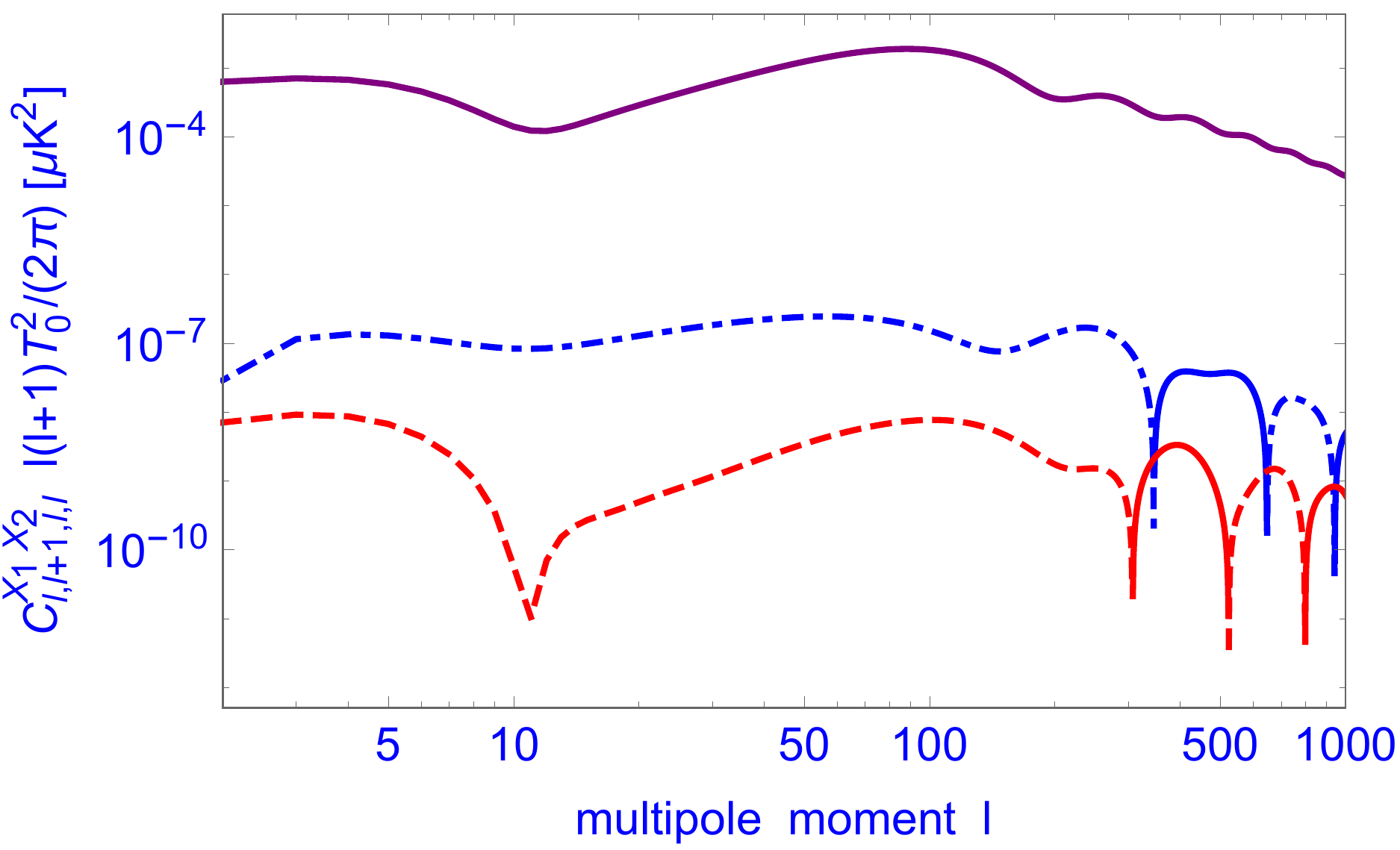}}\\
\caption{The magnitude of $TB$ spectrum $C^{TB}_{l,l+1,l,l}$ (the middle blue dashed-solid curve) and the magnitude of $EB$ spectrum $C^{EB}_{l,l+1,l,l}$ (the bottom red dotted-solid curve), both induced by the cross-correlations, are shown in comparison with the magnitude of the isotropic $BB$ spectrum $C^{BB}_{l,l,0,0}$ (the upper purple dotted curve). 
The left hand side and right hand side figures correspond to the canonical model with $c_s=1$ and the non-canonical model with $c_s=0.1$, respectively. 
Note again that the dotted-dashed (blue) and dashed (red) curves also denote the region where $C^{X_1 X_2}_{l,l+1,l,l}<0$.}
\label{fig2}
\end{center}
\end{figure}

\begin{figure}[hbtp]
\begin{center}
{\includegraphics[height=50mm]{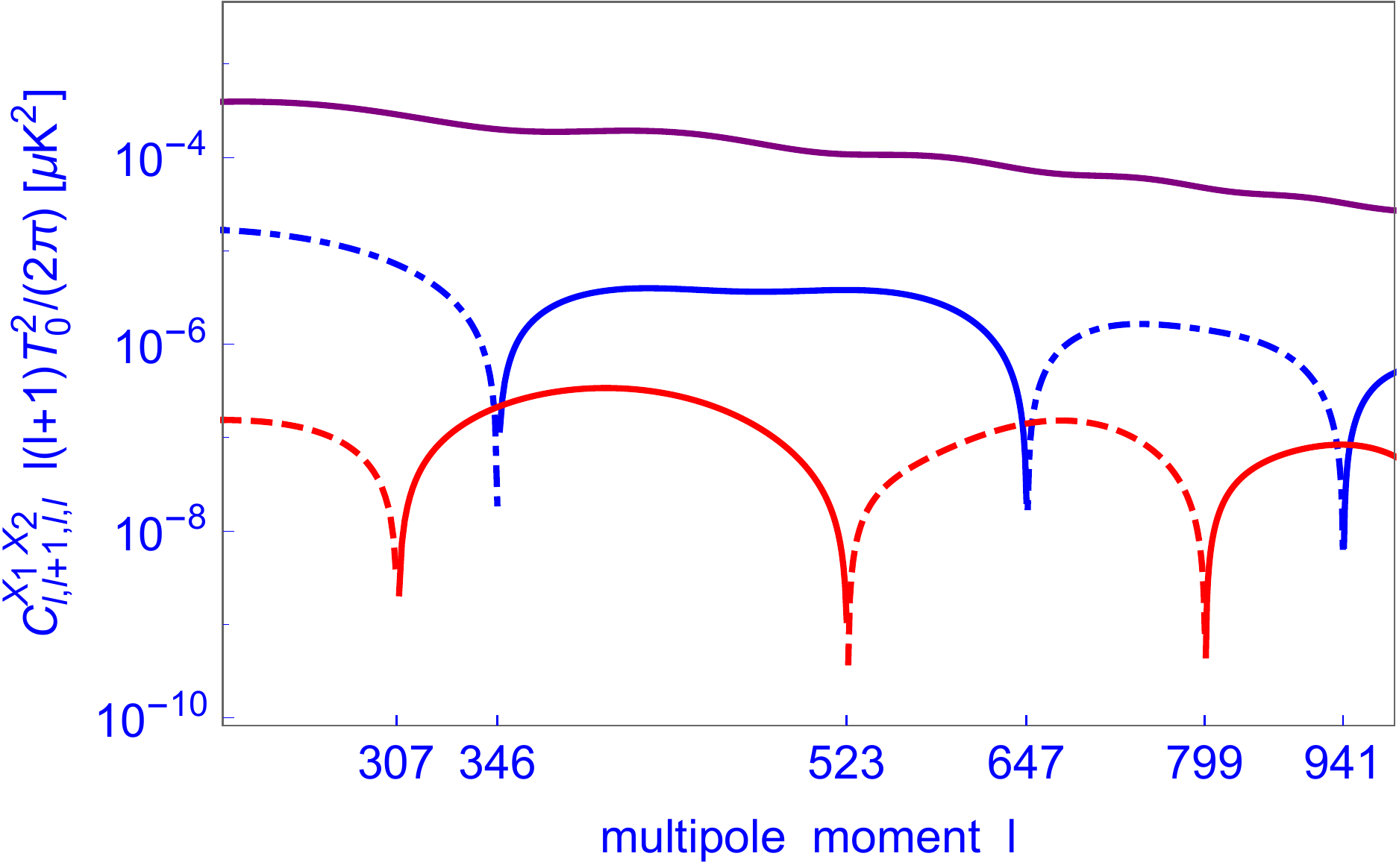}}\quad
{\includegraphics[height=50mm]{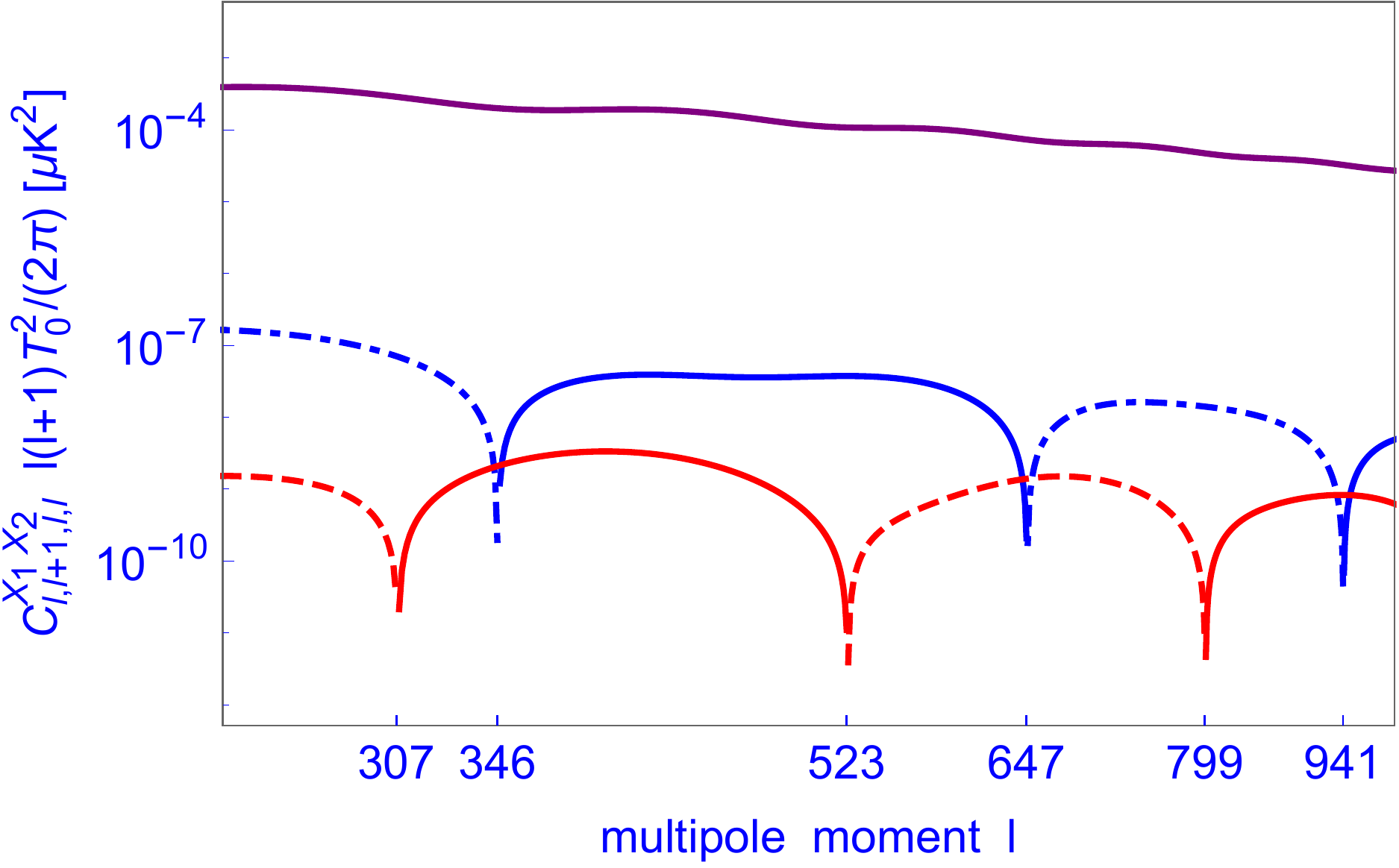}}\\
\caption{Curves shown here are enlarged version of Fig. \ref{fig2}  in order to specify sign-change points of the corresponding spectrum. As a result,  the turning points of $C^{TB}_{l,l+1,l,l}$  and $C^{EB}_{l,l+1,l,l}$ are $l=346,~647, ~941$ and  $l=307,~523, ~799$ respectively.}
\label{fig3}
\end{center}
\end{figure}

Moreover, the TB spectrum $C^{TB}_{l,l+1,l,l}$ (the blue dotted-dashed-solid curve) and the  EB spectrum $C^{EB}_{l,l+1,l,l}$ (the red dotted-solid curve) both induced by the cross-correlation decrease when $c_s$ decreases as shown  in Fig. \ref{fig2}. 
This result can also be verified directly by Eq. (\ref{correlations-cross-2}).

In addition, the points where sign change along the curves shown in Figs. \ref{fig1} and \ref{fig2} can be identified by finding all $l$ where $f(l-1)f(l)<0$. 
Here $f(l)$ denotes the $l$-dependence of the spectrum plotted in Figs. \ref{fig1} and \ref{fig2}.
As a result, we can identify the turning points occur when $l=16$ and $92$ for $C^{TT}_{l,l+2,0,0}$ induced by the anisotropy in the scalar perturbations (blue dashed-solid curve), and $l=48,~164,~315,~617$ and $914$ for $C^{TT}_{l,l+2,0,0}$ induced by the cross-correlations (thick dotted-solid red curves) in Fig. \ref{fig1}.
In addition, the turning points of $C^{TB}_{l,l+1,l,l}$ and $C^{EB}_{l,l+1,l,l}$ are also specified in the partially enlarged graph shown in Fig. \ref{fig3}.


\section{Conclusions}\label{final}
Effect of non-canonical scalar fields on the CMB imprints of the anisotropic inflation has been discussed in details in this article. 
In particular, by using the Bunch-Davies (BD) vacuum state for the non-canonical scalar field \cite{CHKS}, we have obtained a general formalism of tensor-to-scalar ratio as well as a general formalism of the angular power spectra in the scalar perturbations, tensor perturbations, cross-correlations, and linear polarization for the anisotropic inflationary model with non-canonical scalar field. 
These results are shown in Eqs. (\ref{general r-1}), (\ref{off-diagonal-scalar}),  (\ref{off-diagonal-tensor}), (\ref{correlations-cross-1}), (\ref{correlations-cross-2}), (\ref{correlations-linear-1}), and (\ref{correlations-linear-2}).

In order to test the non-canonical anisotropic inflationary models, some most significant spectra have also been plotted numerically with input parameters adopted from recent observations of the Planck satellite,  BICEP2 and Keck array detectors. 
As a result, the magnitude of all spectra in this paper is shown to be much lower than the results shown in Ref. \cite{Imprint2}.
This is derived from the fact that the $r$ and $g_\ast^0$ chosen in this paper are all of ${\cal O}(10^{-2})$. 
In addition, we have shown that the $TT$ spectrum induced by the anisotropy in the scalar perturbations is not the most sensitive among four $TT$ spectra shown in this paper when $c_s <1$. 
This result is also compared with the results for the canonical scalar field with $c_s=1$ in which the magnitude of the $TT$ spectrum induced by the anisotropy in the scalar perturbations turns out to be the largest. 
This is the most distinguishable difference between the canonical and non-canonical anisotropic inflationary models. 
Hopefully, these predictions can be tested in the near future by a more sensitive  primordial gravitational wave observations.
\begin{acknowledgments}
We would like to thank referees very much their comments and suggestions, which are very useful to improve this paper.  T.Q.D. is supported by the Vietnam National Foundation for Science and Technology Development (NAFOSTED) under Grant~No.~103.01-2017.12. W.F.K. is supported in part by the Ministry of Science and Technology (MOST) of Taiwan under Contract No. MOST 108-2112-M-009-002. T.Q.D. would like to thank Prof. Yi Wang very much for sharing some useful parts of his numerical codes used to plot the spectra of anisotropic inflation. T.Q.D. would also like to thank Prof. Julien Lesgourgues very much for his fruitful advice on his numerical codes (CLASS). 
\end{acknowledgments}
\appendix 
\section{The second order correction to the full scalar power spectrum} \label{app1}
In order to derive Eq. (\ref{scalar-power-spectra-2}), we need to perform  the following integration: 
\begin{align} \label{important_integration}
& -\int^\eta_{\eta_{min,1}}d\eta_1 \int^{\eta_1}_{\eta_{min,2}}d\eta_2\langle 0| \left[ \left[\hat{\zeta}^{(0)}_{\text{nc}}(k,\eta) \hat{\zeta}^{(0)}_{\text{nc}}(k',\eta), {H}_{\zeta} (\eta_1) \right],{H}_{\zeta} (\eta_2) \right]|0\rangle \nonumber\\
=& - \frac{16 E_x^2}{H^8}\int^\eta_{\eta_{min,1}}d\eta_1 \int^{\eta_1}_{\eta_{min,2}}d\eta_2 \langle 0| \int d^3k_1\int d^3k_2 \nonumber\\
&\times \frac{1}{ \eta_1^4 \eta_2^4} \left[ \hat{\zeta}^{(0)}_{\text{nc}}(k,\eta) \hat{\zeta}^{(0)}_{\text{nc}}(k',\eta) \hat{\zeta}^{(0)}_{\text{nc}}(k_1,\eta_1)-\hat{\zeta}^{(0)}_{\text{nc}}(k_1,\eta_1) \hat{\zeta}^{(0)}_{\text{nc}}(k,\eta) \hat{\zeta}^{(0)}_{\text{nc}}(k',\eta),\hat{\zeta}^{(0)}_{\text{nc}}(k_2,\eta_2) \right]\delta{\cal E}_x({\bf k_1},\eta_1)\delta{\cal E}_x({\bf k_2},\eta_2)|0\rangle \nonumber\\
\simeq &~\frac{ 4c_s^4 E_x^2  }{k^3 \epsilon^2 M_p^4}\delta^3({\bf k}+{\bf k'})\sin^2\theta \int^\eta_{-1/(c_sk)}d\eta_1 \frac{(\eta^3-\eta_1^3 )}{\eta_1^4 } \int^{\eta_1}_{-1/(c_sk)}d\eta_2 \frac{({\eta^3}-\eta_2^3 )}{\eta_2^4 } \nonumber\\
\simeq &~\frac{2\pi^2}{k^3}\delta^3({\bf k}+{\bf k}')   \frac{c_s^4 E_x^2 N_{c_s k}^2 }{\pi^2 \epsilon^2 M_p^4} \sin^2 \theta.
\end{align}
Here we have used the relation $ \eta_{min,1}=\eta_{min,2} $ due to the factor $\delta^3({\bf k+k'})$. 
In addition, we have also used the identity $N_{c_sk}\equiv \ln|\eta c_s k|$ and the condition $| \eta c_s k |\ll 1$.
\section{The second order correction to the full tensor power spectrum} \label{app2}
In order to derive the tensor power spectrum correction derived from the coupling term $-1/4 h^2(\phi)F^2$, we need to evaluate the tree-level interacting Hamiltonian for tensor perturbations. 
As a result, three different tensor parts of the tree-level interacting Hamiltonian can be shown to be \cite{Imprint}
\begin{eqnarray} 
H_{h1} &=& a^4 {E_x}\int d^3k \delta {\cal E}_x({\bf k},\eta )\frac{1}{\sqrt{2}}\hat{h}_+( -{\bf k},\eta)\sin^2\theta,\\
 H_{h2} &=& a^4 {E_x}\int d^3k \delta{ {\cal E}_y}( {\bf k},\eta)\frac{-1}{\sqrt{2}}\hat{h}_+(-{\bf k},\eta)\sin\theta \cos\theta,\\
 H_{h3}& =& a^4 {E_x}\int d^3k \delta{ {\cal E}_z}( {\bf k},\eta)\frac{-i}{\sqrt{2}}\hat{h}_\times(-{\bf k},\eta )\sin\theta 
\end{eqnarray}
associated with the corresponding tree-level interacting Lagrangian 
\begin{equation}
L_{int,h} = -a^4 E_x \sum_{i=x,y,z}^{} \delta E_i h_{xi}.
\end{equation} 
In addition, the two-point correlation of tensor perturbation reduces to
\begin{align} \label{two-point-tensor-perturbation-1}
\langle 0 \vert \hat h_{ij}({\bf k}) \hat h_{ij}({\bf k'}) \vert 0\rangle \simeq \langle 0 \vert \hat h_{ij}^{(0)}({\bf k}) \hat h_{ij}^{(0)}({\bf k'}) \vert 0\rangle + \delta \langle 0| \hat{h}_\times({\bf k})\hat{h}_\times({\bf k'}) |0\rangle +\delta \langle 0| \hat{h}_+({\bf k})\hat{h}_+({\bf k}') |0\rangle.
\end{align}
Here the cross terms between $\hat{h}_\times ({\bf k})$  and $\hat{h}_+ ({\bf k})$ in Eq. (\ref{two-point-tensor-perturbation-1}) vanishes,
\begin{equation}
\delta \langle 0| \hat{h}_\times ({\bf k})\hat{h}_+({\bf k}') |0\rangle =\delta \langle 0| \hat{h}_+ ({\bf k})\hat{h}_\times({\bf k}') |0\rangle=0
\end{equation}
due to the normalization condition $e^{s}_{ij}({\bf k})e^{\ast s'}_{ij}({\bf k}) =\delta_{ss'}$.
We also need to evaluate a set of corrections for power spectrum. 
The first integration is  
\begin{align}
& \delta \langle 0| \hat{h}_\times({\bf k})\hat{h}_\times({\bf k'}) |0\rangle  \nonumber\\
= & -   \int^\eta_{\eta_{min,1}}d\eta_1 \int^{\eta_1}_{\eta_{min,2}}d\eta_2\langle 0| \left[ \left[\hat{h}^{(0)}_\times({\bf k},\eta) \hat{h}^{(0)}_\times({\bf k'},\eta), {H}_{h3}(\eta_1) \right],{H}_{h3}(\eta_2) \right] |0\rangle \nonumber\\
\simeq & -\frac{16 E_x^2}{9H^4 M_p^4}\sin^2\theta \int^\eta_{\eta_{min,1}}d\eta_1  \frac{(\eta^3-\eta_1^3)}{\eta_1^4} \int^{\eta_1}_{\eta_{min,2}}d\eta_2   \frac{(\eta^3-\eta_2^3)}{\eta_2^4}   \langle 0| {\delta{\cal E}_z(-{\bf k},\eta_1)\delta{\cal E}_z(-{\bf k}',\eta_2)} |0\rangle \nonumber\\
\simeq &~  \frac{2\pi^2 }{k^3} \delta^3({\bf k+k'})\frac{2E_x^2}{ \pi^2 M_p^4 } N_k^2 \sin^2 \theta .
\end{align}
Here  $\eta_{min}=-1/k$ and $N_{k}\equiv \ln|\eta  k|$. 
Consequently, the '$\times$' mode power spectrum correction can be shown to be
 \begin{equation}
\delta {\cal P}_{h_\times}=\frac{2E_x^2}{ \pi^2  M_p^4 } N_k^2 \sin^2 \theta.
\end{equation}
Similarly,  the $+$ mode correction can be derived from 
\begin{align}
\delta \langle 0| \hat{h}_+({\bf k})\hat{h}_+({\bf k}') |0\rangle =& - \mathop{\sum_{A,B=h1,h2}}\int^\eta_{\eta_{min,1}}d\eta_1 \int^{\eta_1}_{\eta_{min,2}}d\eta_2\langle 0| \left[ \left[\hat{h}^{(0)}_+({\bf k},\eta) \hat{h}^{(0)}_+({\bf k}',\eta), {H}_A(\eta_1) \right],{H}_B(\eta_2) \right]|0\rangle
\nonumber \\
\simeq &~ \frac{16E_x^2}{9H^4 M_p^4} \int^\eta_{\eta_{min,1}}d\eta_1  \frac{(\eta^3-\eta_1^3)}{\eta_1^4} \int^{\eta_1}_{\eta_{min,2}}d\eta_2   \frac{(\eta^3-\eta_2^3)}{\eta_2^4} \nonumber\\
&\times \left[ \langle 0| {\delta{\cal E}_x(-{\bf k},\eta_1)\delta{\cal E}_x(-{\bf k}',\eta_2)} |0\rangle \sin^4 \theta - \langle 0| {\delta{\cal E}_x(-{\bf k},\eta_1)\delta{\cal E}_y(-{\bf k}',\eta_2)} |0\rangle \sin^3 \theta \cos\theta \right. \nonumber\\
&\left.- \langle 0| {\delta{\cal E}_y(-{\bf k},\eta_1)\delta{\cal E}_x(-{\bf k}',\eta_2)} |0\rangle \sin^3 \theta \cos\theta+ \langle 0| {\delta{\cal E}_y(-{\bf k},\eta_1)\delta{\cal E}_y(-{\bf k}',\eta_2)} |0\rangle \sin^2 \theta \cos^2\theta \right]. \nonumber\\
\end{align}
Indeed, with the help of 
\begin{align} \label{correlations-1}
\langle 0| {\delta{\cal E}_x(-{\bf k},\eta_1)\delta{\cal E}_x(-{\bf k}',\eta_2) }|0\rangle &= \frac{9H^4}{ {2}k^3}\sin^2 \theta \delta^3({\bf k+k'}) , \\
\label{correlations-2}
\langle 0| {\delta{\cal E}_y(-{\bf k},\eta_1)\delta{\cal E}_y(-{\bf k}',\eta_2)} |0\rangle &= \frac{9H^4}{ {2}k^3} \cos^2 \theta \delta^3({\bf k+k'}),\\
\label{correlations-3}
\langle 0| {\delta{\cal E}_x(-{\bf k},\eta_1)\delta{\cal E}_y(-{\bf k}',\eta_2)} |0\rangle &=\langle 0| {\delta{\cal E}_y(-{\bf k},\eta_1)\delta{\cal E}_x(-{\bf k}',\eta_2)} |0\rangle = -\frac{9H^4}{ {2}k^3} \sin \theta \cos \theta \delta^3({\bf k+k'}) ,
\end{align}
we can show that
\begin{align}
\delta \langle 0| \hat{h}_+({\bf k})\hat{h}_+({\bf k}') |0\rangle  = &~ \frac{{4}E_x^2}{k^3} \int^\eta_{\eta_{min,1}}d\eta_1  \frac{(\eta^3-\eta_1^3)}{\eta_1^4} \int^{\eta_1}_{\eta_{min,2}}d\eta_2   \frac{(\eta^3-\eta_2^3)}{\eta_2^4} \nonumber\\
&\times \left[  \left(\sin^6 \theta +{2}\sin^4 \theta \cos^2\theta +\sin^2 \theta\cos^4\theta \right)\delta^3({\bf k+k'})\right] \nonumber\\
\simeq &~   \frac{2\pi^2 }{k^3} \delta^3({\bf k+k'})\frac{2E_x^2}{ \pi^2 M_p^4 } N_k^2 \sin^2 \theta .
\end{align}
Hence the  '$+$' mode power spectrum correction becomes
\begin{equation}
\delta {\cal P}_{h_+}=\frac{2E_x^2}{ \pi^2 M_p^4 } N_k^2 \sin^2 \theta .
\end{equation}
This is identical to the result  $\delta {\cal P}_{h_\times}$ consistent with the result shown in Eq. (\ref{full-power-spectrum-tensor}).


\section{ ${\cal P}_{\zeta h_{+}}$ and ${\cal P}^{\text{pol}}_h$} \label{app3}
The cross term ${\cal P}_{\zeta h_{+}}$  can be derived  from the two-point correlation:
\begin{align}
&\langle 0| \hat{\zeta}_{\text{nc}}(k,\eta)  \hat{h}_+({\bf k}') |0\rangle \nonumber\\
\simeq &- \int^\eta_{\eta_{min,1}}d\eta_1 \int^{\eta_1}_{\eta_{min,2}}d\eta_2\langle 0| \left[ \left[ \hat{\zeta}^{(0)}_{\text{nc}}(k,\eta) \hat{h}^{(0)}_+({\bf k}',\eta), {H}_\zeta(\eta_1) \right], {H}_{h1}(\eta_2)+{H}_{h2}(\eta_2) \right]|0\rangle \nonumber\\
=&- \frac{ 2 \sqrt{2} E_x^2}{H^8 M_p^4} \int^\eta_{\eta_{min,1}}d\eta_1 \int^{\eta_1}_{\eta_{min,2}}d\eta_2  \frac{1}{\eta_1^4\eta_2^4} \frac{-4 H^4 c_s^2 }{9\epsilon }(\eta^3-\eta^3_1)(\eta^3-\eta^3_2)  \nonumber\\
& \times\left[\langle 0|\delta{\cal E}_x(-{\bf k},\eta_1)\delta {\cal E}_x(-{\bf k}',\eta_2) |0\rangle \sin^2\theta   -\langle 0|\delta{\cal E}_y(-{\bf k},\eta_1)\delta {\cal E}_x(-{\bf k}',\eta_2) |0\rangle \sin\theta \cos\theta \right].
\end{align}
With the help of Eqs. (\ref{correlations-1}), (\ref{correlations-2}), and (\ref{correlations-3}), it can be shown that
\begin{equation}
\langle 0| \hat{\zeta}_{\text{nc}}(k,\eta)  \hat{h}_+({\bf k}') |0\rangle = -    \frac{2\pi^2 }{k^3}\delta^3({\bf k+k'}) \frac{2 E_x^2 c_s^2}{ \sqrt{2} \pi^2 \epsilon M_p^4  }\sin^2 \theta N_{c_s k}N_k .
\end{equation}
Hence the cross power spectrum is given by Eq. (\ref{P-cross}):
\begin{equation}
{\cal P}_{\zeta h_+ } = - \frac{ 2 E_x^2 c_s^2 N_{c_s k}N_k }{ \sqrt{2}\pi^2 \epsilon M_p^4  }\sin^2 \theta = c_s^3 \sqrt{2} \epsilon g_\ast^0     {\cal P}^{\zeta(0)}_{k,{\text{nc}}} \sin^2 \theta.
\end{equation}
Note that  $\eta_{min}=-1/(c_s k) <-1/k$ for $c_s <1$.

In addition, the linear polarization term is defined as ${\cal P}^{\text{pol}}_h=({\cal P}_{h_+} -{\cal P}_{h_\times })/2$ as shown in Eq. (\ref{gap}).

Note that ${\cal P}_{h_+} \simeq {\cal P}_{h_\times }$ for $0$th, $1$st, and $2$nd order in the expansion of the interacting Hamiltonian.
The leading correction is derived from higher order terms.
In addition, $\langle 0| aaa |0\rangle =\langle 0| a^\dagger aa |0\rangle=\ldots=\langle 0| a^\dagger a^\dagger a^\dagger |0\rangle=0$, we can show that the third order term contributes nothing to ${\cal P}^{\text{pol}}_h$. This means that we have to consider at least the fourth order terms. Firstly, we evaluate the following correlation up to the fourth order
\begin{align} \label{fourth-order-1}
&\langle0| \hat{h}_\times({\bf k})\hat{h}_\times({\bf k'}) |0\rangle^{(4)}\nonumber\\
 =&~  \int^\eta_{\eta_{min,1}}d\eta_1 \int^{\eta_1}_{\eta_{min,2}}d\eta_2  \int^{\eta_2}_{\eta_{min,3}}d\eta_3 \int^{\eta_3}_{\eta_{min,4}}d\eta_4 \nonumber\\
&\times \langle 0| \left[\left[\left[\left[\hat{h}^{(0)}_\times({\bf k},\eta) \hat{h}^{(0)}_\times({\bf k'},\eta), {H}_{h3}(\eta_1) \right],{H}_{h3}(\eta_2) \right],{H}_{h3}(\eta_3) \right],{H}_{h3}(\eta_4) \right]|0\rangle \nonumber\\
\simeq &~ \frac{E_x^4}{4 H^{16}} \sin^4\theta \int^\eta_{\eta_{min,1}}d\eta_1 \int^{\eta_1}_{\eta_{min,2}}d\eta_2  \int^{\eta_2}_{\eta_{min,3}}d\eta_3 \int^{\eta_3}_{\eta_{min,4}}d\eta_4 \langle 0| \int d^3k_1\int d^3k_2\int d^3k_3\int d^3k_4 \nonumber\\
 & \times \left[\left[\left[\left[\hat{h}^{(0)}_\times({\bf k},\eta)  \hat{h}^{(0)}_\times({\bf k'},\eta), \hat{h}^{(0)}_\times({\bf k_1},\eta_1) \right],\hat{h}^{(0)}_\times({\bf k_2},\eta_2) \right],\hat{h}^{(0)}_\times({\bf k_3},\eta_3) \right],\hat{h}^{(0)}_\times({\bf k_4},\eta_4)  \right] \nonumber\\
&\times \frac{1 }{\eta_1^4\eta_2^4\eta_3^4\eta_4^4}  \delta{\cal E}_z(\eta_1, {\bf k_1})\delta{\cal E}_z(\eta_2, {\bf k_2})\delta{\cal E}_z(\eta_3, {\bf k_3})\delta{\cal E}_z(\eta_4, {\bf k_4}) |0\rangle.
\end{align}
In order to deal with the complicate integration, we need the following useful results,
\begin{align}
&\int d^3k_1\int d^3k_2 \left[\left[\hat{h}^{(0)}_\times({\bf k},\eta)  \hat{h}^{(0)}_\times({\bf k'},\eta), \hat{h}^{(0)}_\times({\bf k_1},\eta_1) \right],\hat{h}^{(0)}_\times({\bf k_2},\eta_2) \right] \delta{\cal E}_z({\bf k_1},\eta_1 )\delta{\cal E}_z({\bf k_2},\eta_2 )  \nonumber\\
\simeq &-\frac{16H^4}{9M_p^4} (\eta^3-\eta_2^3)(\eta^3-\eta_1^3)\left[ \delta{\cal E}_z({\bf -k'},\eta_1 )\delta{\cal E}_z({\bf -k},\eta_2)+\delta{\cal E}_z({\bf -k},\eta_1)\delta{\cal E}_z({\bf -k'},\eta_2) \right]\nonumber\\
=& -\frac{16H^8}{\sqrt{k^3k'^3} M_p^4 } (\eta^3-\eta_2^3)(\eta^3-\eta_1^3)  \left[ a_2({\bf -k'})+a_2^\dagger({\bf k'})  \right]\left[ a_2({\bf -k})+a_2^\dagger({\bf k})   \right] ,
\\
& \int d^3k_3\int d^3k_4 \left[\left[1, \hat{h}^{(0)}_\times({\bf k_3},\eta_3)\right],\hat{h}^{(0)}_\times({\bf k_4},\eta_4) \right] \delta{\cal E}_z({\bf k_3},\eta_3)\delta{\cal E}_z({\bf k_4},\eta_4) \nonumber\\
\simeq &-i\frac{4H^2}{3M_p^2} \int d^3k_3 (\eta_3^3-\eta_4^3) \delta{\cal E}_z({\bf k_3},\eta_3 )\delta{\cal E}_z({\bf -k_3},\eta_4)\nonumber\\
\simeq& -i\frac{6H^6}{M_p^2}(\eta_3^3-\eta_4^3) \int \frac{d^3k_3}{ k_3^3 } \left[ a_2({\bf k_3})+a_2^\dagger({\bf -k_3}) \right] \left[ a_2({\bf -k_3})+a_2^\dagger({\bf k_3})\right].
\end{align}
Here  $\delta {\cal E}_i$ is defined in Eq. (\ref{def.of.delta.cal.E}).
As a result, the integration in Eq. (\ref{fourth-order-1}) reduces to
\begin{align}
&\langle0| \hat{h}_\times({\bf k})\hat{h}_\times({\bf k'}) |0\rangle^{(4)} \nonumber\\
 \simeq & ~ \frac{ 24 E_x^4}{ H^{2} M_p^6} \sin^4\theta \int^\eta_{\eta_{min,1}}d\eta_1 \int^{\eta_1}_{\eta_{min,2}}d\eta_2  \int^{\eta_2}_{\eta_{min,3}}d\eta_3 \int^{\eta_3}_{\eta_{min,4}}d\eta_4  \langle 0|  \int  d^3k_3  
\frac{(\eta^3-\eta_1^3)(\eta^3-\eta_2^3)(\eta_3^3-\eta_4^3) }{\eta_1^4\eta_2^4\eta_3^4\eta_4^4}  \frac{i}{\sqrt{k^3k'^3} k_3^3 } \nonumber\\
&\times \left[ a_2({\bf -k'})+a_2^\dagger({\bf k'})   \right]\left[ a_2({\bf -k})+a_2^\dagger({\bf k})   \right] \left[ a_2({\bf k_3})+a_2^\dagger({\bf -k_3})\right]\left[ a_2({\bf -k_3})+a_2^\dagger({\bf k_3}) \right]  |0\rangle \nonumber\\
\simeq &~ \frac{ E_x^4 N_{k}^4}{ H^{2}M_p^6} \sin^4\theta \langle 0|  \int  d^3k_3 \frac{i}{\sqrt{k^3k'^3} k_3^3 }  \left[ a_2({\bf -k'}), a_2^\dagger({\bf k'})   \right] \left[ a_2({\bf k_3})a_2^\dagger({\bf k_3})+a_2^\dagger({\bf k_3}) a_2({\bf k_3}) \right]  |0\rangle.
\end{align}
Note also that 
\begin{equation}
\int  d^3k_3 \frac{i}{\sqrt{k^3k'^3} k_3^3 }  \langle 0| \left[ a_2({\bf -k'}), a_2^\dagger({\bf k'})   \right] \left[ a_2({\bf k_3})a_2^\dagger({\bf k_3})+a_2^\dagger({\bf k_3}) a_2({\bf k_3}) \right]  |0\rangle \sim   \frac{Q_\times}{k^3} \delta^3({\bf k+k' }),
\end{equation}
with $Q_\times \simeq {\cal O}(1)$  an undetermined number.
Hence it can be shown that
\begin{equation}
\langle0| \hat{h}_\times({\bf k})\hat{h}_\times({\bf k'}) |0\rangle^{(4)} \sim \frac{2\pi^2}{k^3} \delta^3({\bf k+k' })   \frac{ Q_\times E_x^4 N_{ k}^4 }{2H^2\pi^2 M_p^6}  \sin^4\theta.
\end{equation} 
This implies that
\begin{equation}
\delta {{\cal P}^{(4)}_{h_\times}}\sim   \frac{ Q_\times E_x^4 N_{ k}^4 }{2H^2\pi^2 M_p^6}  \sin^4\theta.
\end{equation}
Similarly,  $\langle0| \hat{h}_+({\bf k})\hat{h}_+({\bf k'}) |0\rangle^{(4)}$ can be shown to be
\begin{align}
\langle0| \hat{h}_+({\bf k})\hat{h}_+({\bf k'}) |0\rangle^{(4)} & \simeq \frac{ E_x^4 N_{k}^4}{ H^{2}} \sin^4\theta \langle 0|  \int  d^3k_3 \frac{i}{\sqrt{k^3k'^3} k_3^3 }  \left[ a_1({\bf -k'}), a_1^\dagger({\bf k'})   \right] \left[ a_1({\bf k_3})a_1^\dagger({\bf k_3})+a_1^\dagger({\bf k_3}) a_1({\bf k_3}) \right]  |0\rangle \nonumber\\
&\sim \frac{2\pi^2}{k^3}\delta^3({\bf k+k' })  \frac{ Q_+ E_x^4 N_{ k}^4 }{2H^2\pi^2 M_p^6}  \sin^4\theta ,
\end{align}
with $Q_+ \simeq {\cal O}(1)$ another undetermined number. 
Note that $Q_+ \ne Q_\times$ for $a_1({\bf k_3}) \neq a_2({\bf k_3})$.
Consequently, $\delta {{\cal P}^{(4)}_{h_+}}$ can be shown to be
\begin{equation}
\delta {{\cal P}^{(4)}_{h_+}}\simeq   \frac{ Q_+ E_x^4 N_{ k}^4 }{2H^2\pi^2 M_p^6}  \sin^4\theta.
\end{equation}
As a result, ${\cal P}^{\text{pol}}_h$ defined in \cite{Imprint2} reduces to
\begin{align}
{\cal P}^{\text{pol}}_h \sim \frac{1}{2}\left(\delta {{\cal P}^{(4)}_{h_+}}-\delta {{\cal P}^{(4)}_{h_\times}}\right) \sim  \frac{ E_x^4 N_{ k}^4 }{4H^2\pi^2 M_p^6}  \sin^4\theta \left(Q_+ - Q_\times\right) \sim g_l   {\cal P}^{h(0)}_{k,\text{nc}} \sin^4 \theta,
\end{align}
with  
\begin{equation}
g_l \sim  {\cal O} \left(\frac{\epsilon^2 (g_\ast^0)^2}{16}\right) .
\end{equation}
\section{Spin-$i$-weighted spherical harmonics} \label{app4}
The definition and some of the properties of the spin-$i$-weighted spherical harmonics, the $_i Y_{lm} (\theta,\phi)$, will be listed in this section for reference. 
Indeed, the definition of $_i Y_{lm} (\theta,\phi)$ is given by ~\cite{WHMW}
\begin{eqnarray}
_s Y_{lm}(\theta,\phi) = &&~ \left(\frac{2l+1}{4\pi}\right)^{1/2} {\cal D}^l_{-s,m}(\theta,\phi,0)\\
= &&~\left[\frac{2l+1}{4\pi} \frac{(l+m)! (l-m)!}{(l+s)! (l-s)!}\right]^{1/2}\left[\sin\left(\frac{\theta}{2}\right)\right]^{2l} \nonumber\\
&& ~ \times \sum\limits_r \left(
\begin{array}{c}
 l-s \\
 r
\end{array}
\right) \left(
\begin{array}{c}
 l+s \\
 r+s-m
\end{array}
\right) (-1)^{l-r-s} e^{im\phi}\left[\cot \left(\frac{\theta}{2}\right)\right]^{2r+s-m}.
\end{eqnarray}
Here
\begin{eqnarray}
{\cal D}^l_{-s,m}(\theta,\phi,\psi)=\sqrt{\frac{4\pi}{2l+1}} {_s Y_{lm}}(\theta,\phi) e^{-is\psi},
\end{eqnarray}
denotes the rotation matrix with the Euler angles ($\phi,\theta,\psi$). 
Furthermore, some useful properties of $_i Y_{lm} (\theta,\phi)$ are listed below ~\cite{WHMW}
\begin{eqnarray}
_i Y_{lm}^\ast (\theta,\phi) =&&~ {(-1)^{m+i}} {_{-i} Y_{l,-m}} (\theta,\phi) , \\
\int{\left(_i Y_{lm}^\ast\right) \left({_i Y_{l'm'}}\right) d\Omega} =&&~ \delta_{ll'} \delta_{mm'},\\
\sum\limits_{l,m} \left[_i Y_{lm}^\ast (\theta,\phi) \right]\left[_i Y_{lm}(\theta',\phi') \right] = &&~ \delta(\phi-\phi')\delta(\cos\theta -\cos \theta'),\\
_i Y_{lm} \to && ~ (-1)^l {_{-i} Y_{lm}}, \\
\left(_{i_1} Y_{l_1m_1}\right) \left({_{i_2} Y_{l_2 m_2}}\right) = &&~ \frac{\sqrt{(2l_1+1)(2l_2+1)}}{4\pi} \nonumber\\
&& ~\times \sum\limits_{l,m,i}\langle l_1,l_2;m_1,m_2| l_1,l_2;l,m\rangle \langle l_1,l_2;-i_1,-i_2| l_1,l_2;l,-i\rangle  \sqrt{\frac{4\pi}{2l+1}} \left({_i Y_{lm}}\right),
\end{eqnarray}
with $\langle{~} |{~} \rangle$ the Clebsch-Gordan coefficients \cite{CG}.

\end{document}